\documentclass[aps,amsmath,amssymb,aps,prd,superscriptaddress,groupedaddress,nofootinbib]{revtex4}  
\usepackage{graphics}
\usepackage{epsfig}
\usepackage{xcolor}

\newcommand{\nc}{\newcommand}
\nc{\beq}{\begin{equation}}
\nc{\eeq}{\end{equation}}
\nc{\beqa}{\begin{eqnarray}}
\nc{\eeqa}{\end{eqnarray}}

\def\bc{\begin{center}}
\def\ec{\end{center}}

\def\to{\rightarrow}

\def\gsim{\mathrel{\mathpalette\atversim>}}

\def\bc{\begin{center}}
\def\ec{\end{center}}

\def\MPL{\overline{M}_{PL}}

\def\gsim{\mathrel{\rlap{\lower4pt\hbox{\hskip1pt$\sim$}}

    \raise1pt\hbox{$>$}}}       %greater than or approx. symbol

\def\gsim{\mathrel{\rlap{\lower4pt\hbox{\hskip1pt$\sim$}}
    \raise1pt\hbox{$>$}}}       %greater than or approx. symbol

\usepackage{multirow}

\begin{document}
\title{Quantum Black Holes and their Lepton Signatures at the LHC
with CalCHEP}
\author{Alexander Belyaev}
\email{a.belyaev@soton.ac.uk}
\affiliation{School of Physics and
Astronomy,
University of Southampton, Highfield, Southampton SO17 1BJ, UK}
\affiliation{Particle Physics Department, Rutherford Appleton Laboratory,
Chilton, Didcot, Oxon OX11 0QX, UK.}
\author{Xavier Calmet}
\email{x.calmet@sussex.ac.uk}
\affiliation{Physics and Astronomy, 
University of Sussex, Falmer, Brighton, BN1 9QH, UK }

\begin{abstract}
We discuss a field theoretical framework to describe the interactions of non-thermal quantum black holes (QBHs) with particles of the Standard Model. We propose a non-local Lagrangian to describe the production of these QBHs  which is designed  to reproduce the geometrical cross section $\pi r_s^2$ for black hole production where $r_s$ is the Schwarzschild radius.
This model is implemented into CalcHEP package and is publicly available at the High Energy Model Database (HEPMDB) for simulation of QBH events at the LHC and future colliders.
We present the first phenomenological application of the QBH@HEPMDB
model with spin-0 neutral QBH giving rise the
 $e^+e^-$ and $e\mu$  signatures at the LHC@8TeV and LHC@13TeV and
produce the respective projections for the LHC
in terms of limits on the reduced Planck mass, $\overline{M}_{PL}$ and the number of the
extra-dimensions $n$.
\end{abstract}  
\maketitle

%\pacs{}

%%%%%%%%%%%%%%%%%%%%%%%%%%%%%%%%%%%%%%%%%%%%%%%%%%%%%%%%%%%%%%%%
%%%
%%%                     INTRODUCTION
%%%
%%%%%%%%%%%%%%%%%%%%%%%%%%%%%%%%%%%%%%%%%%%%%%%%%%%%%%%%%%%%%%%%
%xc editing
\newpage

\section{Introduction and model setup}

Realising that the Planck mass \cite{ArkaniHamed:1998rs,Randall:1999ee,Calmet:2008tn,Calmet:2010nt,Calmet:2014gya}  may not be around $10^{19}$ GeV but that it is a model dependent quantity which could even be as low as a few TeVs  has had a considerable impact in particle physics. One of the most amazing signatures of these models would be the creation of small black holes at colliders \cite{Dimopoulos:2001hw,Banks:1999gd,Giddings:2001bu,Dai:2007ki,Calmet:2008dg} or in the scattering of cosmic rays  \cite{Feng:2001ib,Anchordoqui:2003ug,Anchordoqui:2001cg,Anchordoqui:2003jr,Calmet:2008rv,Calmet:2012mf,Arsene:2013nca,calmetbook1} in the upper atmosphere of our planet. While it is now well appreciated that semi-classical black holes cannot be produced at the LHC because this collider is not energetic enough even if the Planck mass is at a few TeVs \cite{Meade:2007sz}, the possibility remains to produce non-thermal quantum black holes \cite{Calmet:2008dg}. 

While the semi-classical black holes, which have been extensively studied  \cite{Eardley:2002re,D'Eath:1992hb,Hsu:2002bd,Dimopoulos:2001hw,Banks:1999gd,Giddings:2001bu,Feng:2001ib,Anchordoqui:2003ug,Anchordoqui:2001cg,Anchordoqui:2003jr,Dai:2007ki,calmetbook1},  have masses between 5 to 20 times larger than the Planck scale and are thus thermal objects, most quantum black holes are those with masses of the order of the Planck scale. They will thus be non-thermal objects and their decomposition is thus not expected to be well described by Hawking radiation. They will rather explode into just  a few particles resembling strong gravitational re-scattering. It is thus tempting to treat them as particles with a mass and carrying the quantum numbers of the particles which created them. In a proton collider, there will be quarks and gluons.  The aim of this paper is to extend the work presented in \cite{Calmet:2011ta} where a field theoretical model to describe the interactions of non-thermal quantum black holes with the particles of the standard model was proposed. This framework assumes that quantum black holes can be treated as quantum fields, i.e. they are classified according to representations of the Lorentz group. Furthermore, they are classified according to their transformations under the gauge groups of the standard model. This fixes their interactions with matter.

In a proton-proton collider, quantum black holes would be produced from the collisions of quarks and gluons if the Planck scale is low enough. We are thus particularly interested in the quantum black holes carrying QCD and QED quantum numbers and with spins 0,1/2 and 1 since these should be the lowest lying states. We shall treat the mass spectrum as being discrete. Generically speaking, quantum black holes form representations of SU(3)$_c$ and carry a QED charge. The process of two partons $p_i$, $p_j$ forming a quantum black hole in the $c$ representation of SU(3)$_c$ and charge $q$ as: $p_i+p_j \to$ QBH$_c^q$ is considered in  \cite{Calmet:2008dg}. The following different transitions are possible at a proton collider:
\begin{itemize}
\item[a)] ${\bf 3} \times {\bf \overline 3}= {\bf 8} + {\bf 1}$ \\
\item[b)] ${\bf 3} \times {\bf 3}= {\bf 6} + {\bf \overline 3}$\\
\item[c)] ${\bf 3} \times {\bf 8}= {\bf 3} + {\bf \overline 6}+ {\bf 15}$\\
\item[d)] ${\bf 8} \times {\bf 8}= {\bf 1}_S + {\bf 8}_S+ {\bf 8}_A+{\bf 10} + {\bf \overline{10}}_A+ {\bf 27}_S$
\end{itemize}
Most of the time the black holes which are created carry a SU(3)$_c$ charge and come in different representations of SU(3)$_c$ as well as QED charges. This allows the prediction of  how they will be produced or decay. The aim of this work is to propose a framework to describe these interactions. We shall assume that the cross section for quantum black holes can be extrapolated from the classical and semi-classical cases, i.e. that it is given by the geometrical formula (see e.g. \cite{Eardley:2002re,D'Eath:1992hb})
\begin{eqnarray}
\sigma=\pi r_{s}^{2}
\end{eqnarray}
where $r_{s}$ is the four-dimensional Schwarzschild radius
\begin{eqnarray}
r_{s}(s,\overline{M}_{PL})=\frac{\sqrt{s}}{4\pi\overline{M}_{PL}^{2}},
\end{eqnarray}
where $s$ is the invariant mass of the colliding  particles,
which upon  exceeding the reduced Planck mass, $\overline{M}_{PL}$
creates the respective QBH.
Therefore, in terms of  $s$ and $\overline{M}_{PL}$, the cross section of the QBH production takes a form
\begin{equation}\label{eq:qbh_cs}
\sigma=\frac{1}{16\pi}\frac{s}{\overline{M}_{PL}^4}\Theta(\sqrt{s}-\overline{M}_{PL})
\end{equation}
where we have assumed that the threshold mass for quantum black holes is identified with the Planck mass. Note that the quantum black holes described here are assumed to have a continuous mass spectrum despite some indications that their mass spectrum could be discrete \cite{Calmet:2012fv}. Since we are considering a continuous mass spectrum we have to assume that quantum black hole couplings to long wavelength and highly off-shell perturbative modes are suppressed \cite{Calmet:2008dg}. Otherwise their contribution to low energy observables such as $K_L$ decays would have been noticed a long time ago. Note that there are no such constraints if the mass spectrum of quantum black holes is indeed discrete \cite{calmetbook2}.

We shall first reconsider the production of spinless quantum black holes in the collisions of two fermions (quarks for example with the appropriate colour factor). We start with the Lagrangian
\begin{equation}\label{eq:L1}
L_{fermion+fermion}=\frac{c}{\overline{M}_{PL}^{2}}\partial_{\mu}\partial^{\mu}\phi\bar{\psi}_{1}\psi_{2}+h.c. 
\end{equation}
where $c$ is the (non-local) parameter we will use to match the semiclassical cross section, $\overline{M}_{PL}$ is the reduced Planck mass, $\phi$ is a scalar field representing the quantum black hole, and $\psi_{i}$ is a fermion field. The cross section for $\phi$ production is:
\begin{eqnarray}
\sigma(2\psi\rightarrow\phi)=\frac{\pi}{s}\left|\textit{A}\right|^{2}\delta(s-M^{2}_{BH})
\end{eqnarray}
where $M_{BH}$ is the mass of the black hole, $s=(p_{1}+p_{2})^{2}$ and $p_{1}$,$p_{2}$ are the four-momenta of $\psi_{1}$ $\psi_{2}$. We find
\begin{eqnarray}
\left|\textit{A}\right|^{2}=s^{2}\frac{c^{2}}{\overline{M}_{PL}^{4}}[s-(m_{1}+m_{2})^{2}]
\end{eqnarray}
where $m_{1}$ and $m_{2}$ are the masses of the fermions $\psi_{1}$ and $\psi_{2}$.
We now compare this cross section with the geometrical cross section. If we use the representation for the delta-function
written in the form of the Poisson kernel,
\begin{eqnarray}
\delta(s-M_{BH}^{2})=\frac{\Gamma M_{BH}}{\pi [(s-M_{BH}^2)^{2}+\Gamma^2 M_{BH}^2]}
\end{eqnarray}
where $\Gamma$ is the decay width of $\phi$, we find:

\begin{equation}\label{eq:cc}
c^{2}=\frac{9}{4}\frac{4s^{\frac{3}{2}}-8sM_{BH}+4\sqrt{s}M_{BH}^{2}
+\sqrt{s}\Gamma^{2}}{\Gamma\pi[s-(m_{1}+m_{2})^{2}]}
\end{equation}

Finally $\Gamma$ can be calculated using the Lagrangian (\ref{eq:L1}) as:
\begin{equation}
\Gamma=\frac{c^{2}}{8 \pi}   \frac{M_{BH}\sqrt{(M_{BH}^2-(m_1+m_2)^2)(M_{BH}^2-(m_1-m_2)^2)}}{\overline{M}_{PL}^4}
\end{equation}
We can thus find an expression for our non-local parameter $c$ by inserting $\Gamma$ into the expression for $c$ (\ref{eq:cc}). In the case $m_1=m_2=0$, one has a remarkably simple expression:
\begin{equation}
c^2=\frac{8 \pi \overline{M}_{PL}^4 (s-M_{BH}^2)}{M_{BH}^3 \sqrt{128 \pi^2 \overline{M}_{PL}^4 s -M_{BH}^6}}
\end{equation}

One can see that non-local behaviour of the $c$-coupling is quite non-trivial -- it actually compensates the
Breit-Wigner behaviour of the squared matrix element which would appear in case of constant $c$
and leads to the expected $s$-dependence of the cross section  from Eq.(\ref{eq:qbh_cs}).
It is important to realize that the QBH does not appear as a resonance since its parton level cross section
is constructed to reproduce the semi-classical cross section as a function of $s$. With this in mind,
we have  also found an alternative approach to the  construction of the
QBH Lagrangian. If we recall that the Lagrangian for the four-fermion interactions
\begin{equation}
{\cal L}_{cont}=
\frac{g_c}{\Lambda^2}\bar{\psi}_i\psi_i\bar{\psi}_j\psi_j
\end{equation}
provides the squared matrix element for $\bar{\psi}_i\psi_i \to \bar{\psi}_j\psi_j$
scattering, 
\begin{equation}
|M|^2=\frac{g_c^2}{\Lambda^4}
\end{equation}
and the respective total $\bar{\psi}_i\psi_i \to \bar{\psi}_j\psi_j$ cross section
(neglecting the fermion masses)
\begin{equation}
\sigma=\frac{1}{16 \pi s} |M|^2 = \frac{g_c^2}{16 \pi}\frac{s}{ \Lambda^4},
\end{equation}
then after comparison with Eq.(\ref{eq:qbh_cs}), we can immediately see
that  Eq.(\ref{eq:qbh_cs}) can be reproduced by 4-fermion interactions
with $g_c=1$ and $\Lambda=\overline{M}_{PL}$. In case of different  fermion
species involved in $2\to 2$ process of the QBH production and decay, 
$g_c$ will include the respective number of degrees of freedom 
to correctly reproduce the QBH branching fractions.
In the scenario under study we consider the case of spin-0 neutral QBH production
which preserves $SU(3)\times SU(2)\times U(1)$ gauge invariance but does not conserve flavour. We wish to emphasise that we are here only considering colourless quantum black holes which explains why our branching ratios are different from those of \cite{Calmet:2008tn} (see also \cite{Gingrich:2009hj}).
We have found that the most elegant and practical way to express these 
contact interactions is to use the auxiliary, non propagating scalar field, $X$
which enters the following Lagrangian
\begin{equation}\label{eq:ci}
{\cal L}_{cont}^{X}=
g_c \left(
\sum_{leptons} \bar{\psi}_i^\ell\psi_j^\ell X
 + 
 \sum_{quarks} \bar{\psi}_i^q\psi_j^q X\right),
\end{equation}
where $i,j$ are lepton and quark flavour indices, propagator of ``contact" 
$X$ field is $\frac{i}{M_{PL}^2}$, $g_c=(n_l+ 3 n_q)^{-1/4}$ is the normalisation factor
accounting number of lepton, $n_l=9$ and quark, $n_q=18$ combinations, 
including the  quark colour factor.
This Lagrangian exactly reproduces the  cross section  of the 
spinless neutral QBH production and decay in the $2\to 2 $ fermion process as a function of $s$ at the parton level
(up to the multiplicative  trivial form-factor $FF = \Theta(\sqrt{s}-\overline{M}_{PL})$).

Our results could be generalised easily to the case of initial state particles with different spins and colours for which the approach of the contact interactions also works successfully
as one can check using dimensional analysis approach.

The result can be also generalised to 
the case of higher dimensional quantum black holes for which the
Schwarzschild radius is given by  (see e.g. \cite{Dimopoulos:2001hw})
\begin{eqnarray}
r_s(s,n,\overline{M}_{PL}) = k(n) \overline{M}_{PL}^{-1} (\sqrt{s}/\overline{M}_{PL})^{1/(1+n)}
\end{eqnarray}
where $n$ is the number of extra-dimensions, $\overline{M}_{PL}$ the 4+n reduced Planck mass and $k(n)$ reads
\begin{eqnarray}
k(n) =\left( 2^n \sqrt{\pi}^{(n-3)} \frac{\Gamma((3+n)/2)}{2+n} \right)^{1/(1+n)}.
\end{eqnarray}
The respective form factors  for the case of $n$-dimensions
which should be introduced for the parton-level cross section to reproduce the correct cross section
from the Lagrangian with contact interactions (\ref{eq:ci}) is 
\begin{equation}\label{eq:ff}
FF= \left(4 \pi k(n)\right)^2\left(\frac{\overline{M}_{PL}}{\sqrt{s}}\right)^\frac{2n}{1+n} \Theta(\sqrt{s}-\overline{M}_{PL}).
\end{equation}
Here the case $n=0$ corresponds to 4-dimensional models with low scale quantum gravity \cite{Calmet:2008tn,Calmet:2010nt,Calmet:2014gya}, $n=1$ to Randall Sundrum \cite{Randall:1999ee} brane world model\footnote{Note that will treat the Randall Sundrum quantum black holes as ADD ones with $n=1$. While the cross section for semi-classical black holes in the case of RS differs from that obtained using the Schwarzschild metric \cite{Meade:2007sz}, this is an unnecessary refinement for quantum black holes whose quantum geometry is anyway very poorly understood.} and $n\ge 2$ to ADD model \cite{ArkaniHamed:1998rs}. Note that there are astrophysical constraints on $n=2,3,4$ ADD which shift exclude a Planck mass in the few TeV region, it is however interesting to consider bounds from QBHs which are independent of those coming from Kaluza Klein modes which lead to the astrophysical constraints.
%%%%%%%%%%%%%%%%%%%%%
%%%%%%%%%%%%%%%%%%%%%%
%%%%%%%%%%%%%%%%%%%%%
%%%%%%%%%%%%%%%%%%%%%%
\section{Phenomenology of the QBHs at the LHC}

To study the phenomenology of the QBHs production we have implemented  interactions given by Eq.(\ref{eq:ci}-\ref{eq:ff})
into CalcHEP software package~\cite{Belyaev:2012qa} 
as a QBH model which is publicly available at the High Energy Physics Model Database (HEPMDB)~\cite{hepmdb} under the link \url{http://hepmdb.soton.ac.uk/hepmdb:1113.0146}.
We also would like to note an important feature of CalcHEP which allows the implementation of non-trivial 
form factors at the user level which was one of the key points in the implementation of this model.
We emphasise that the Lagrangian we are proposing to describe the interactions of quantum black holes with particles of the Standard Model should not be regarded as an effective theory in the usual sense, it is rather an effective manner to describe the interactions of these black holes with usual particles.

This model is publicly available at HEPMDB which  provides HEP community 
with a new QBH Monte-Carlo (MC) generator (QBH@HEPMDB), and is an alternative to existing  BlackMax~\cite{Dai:2007ki} and QBH~\cite{Gingrich:2009da} MC generators.
We would like to stress that QBH@HEPMDB model is available for download  and
allows (at HEPMDB website or using CalcHEP locally) to  evaluate cross sections and
generate parton-level events in generic  Les Houches Event (LHE) format~\cite{Alwall:2006yp} which can be {\it independently} used in subsequent analysis
using various general purpose MC generators and detector simulation software.
In this paper we present the first phenomenological application of  the QBH@HEPMDB model with spin-0 neutral QBH [QBH(0,0)]
to $e^+e+$ and $e\mu$ signature  at the LHC@8TeV and LHC@13TeV.  We produce the respective projections for the LHC to probe QBH parameter space.
The model can be easily extended for QBHs with other charges and spins using the same approach as described above.
In our calculations we have used CTEQ6L\cite{Pumplin:2002vw}  parameterisation for the 
parton density functions (PDFs) while the QCD scale was fixed to $\overline{M}_{PL}$.
The parameter space of the model under study is the reduced Planck mass, $\overline{M}_{PL}$,
which sets the threshold for the QBH production as well the number of the
extra-dimensions $n$.

\begin{figure}[htb]
\epsfig{width=0.5\textwidth,file=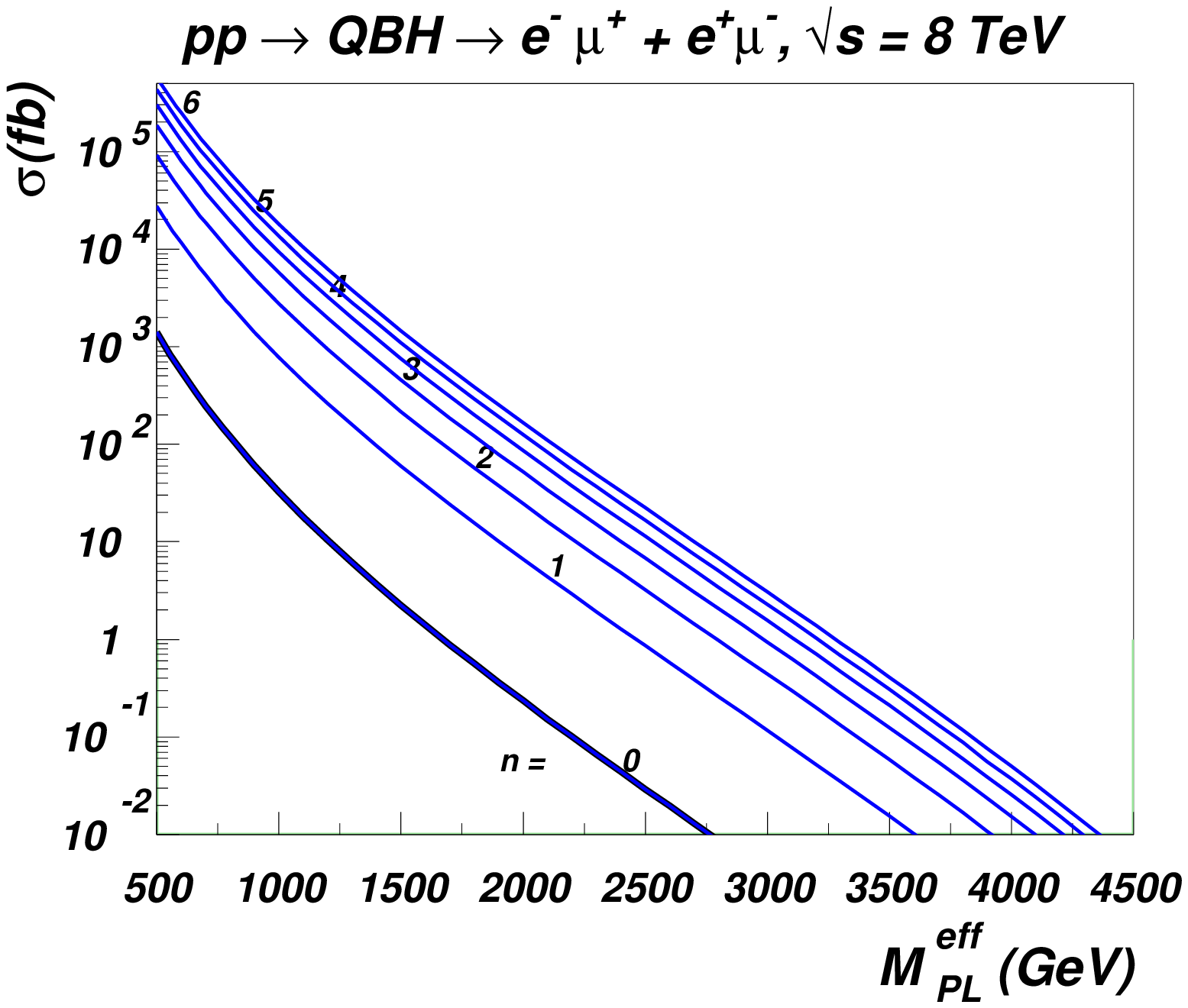}%
\epsfig{width=0.5\textwidth,file=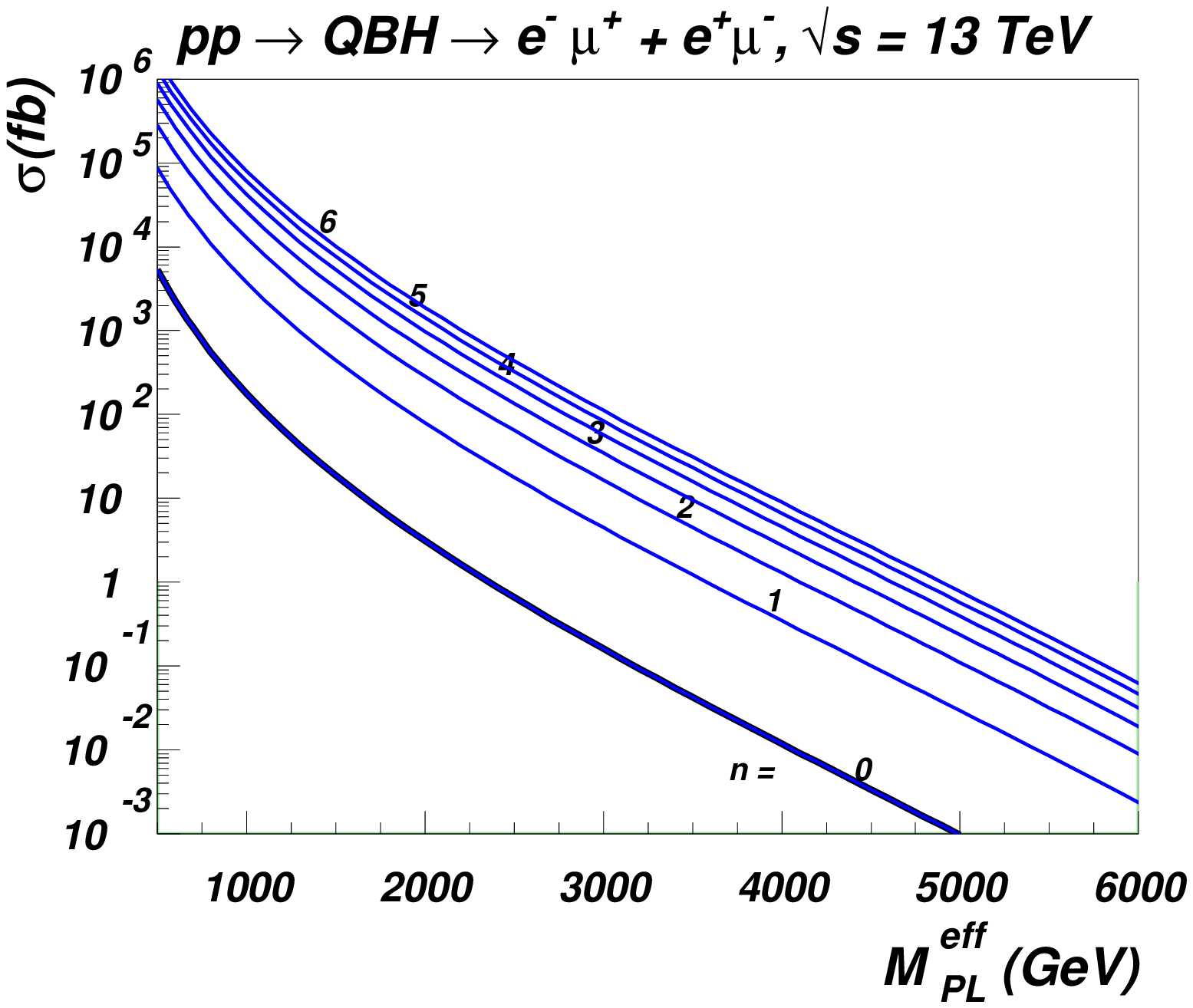}\\
\epsfig{width=0.5\textwidth,file=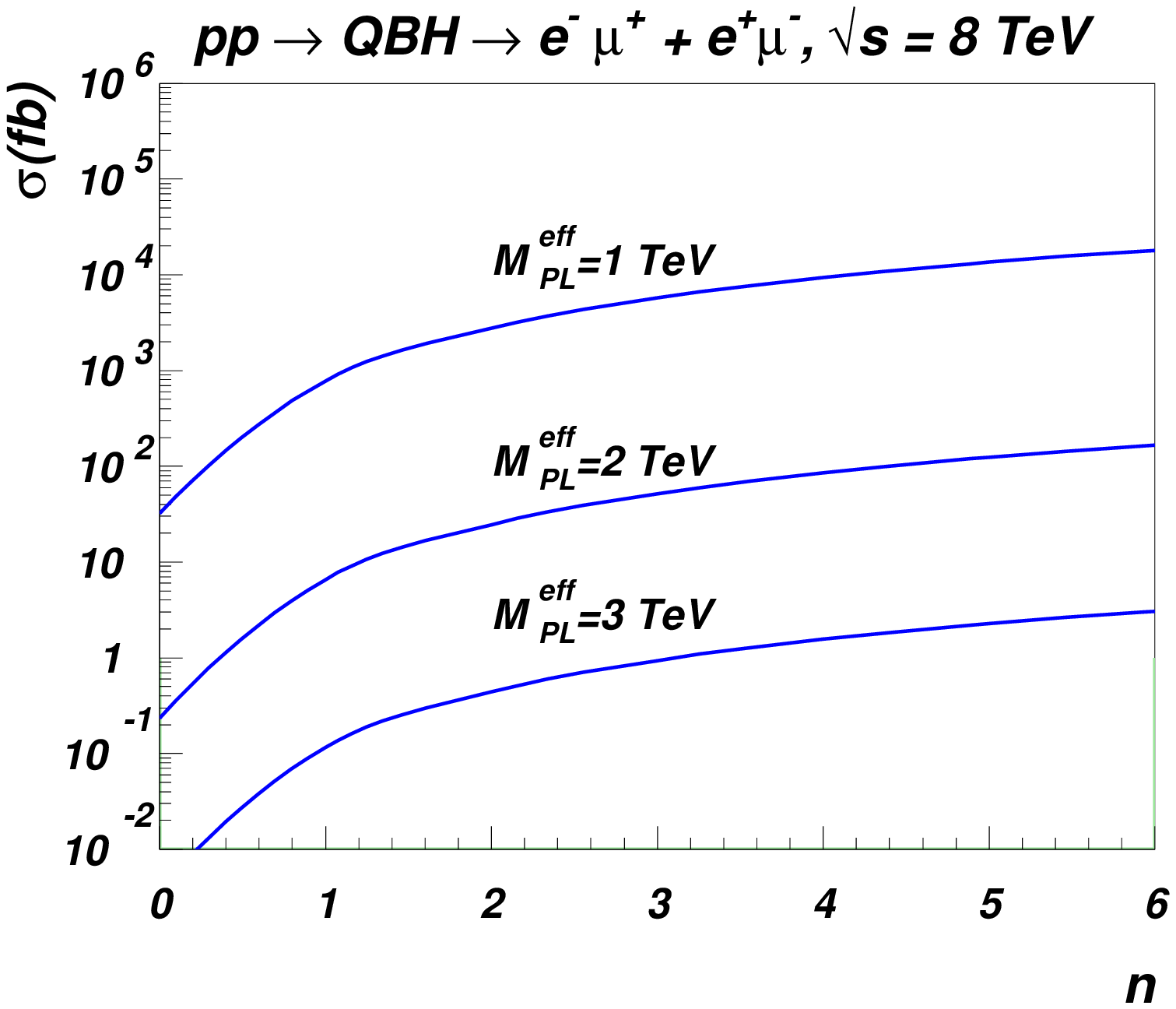}%
\epsfig{width=0.5\textwidth,file=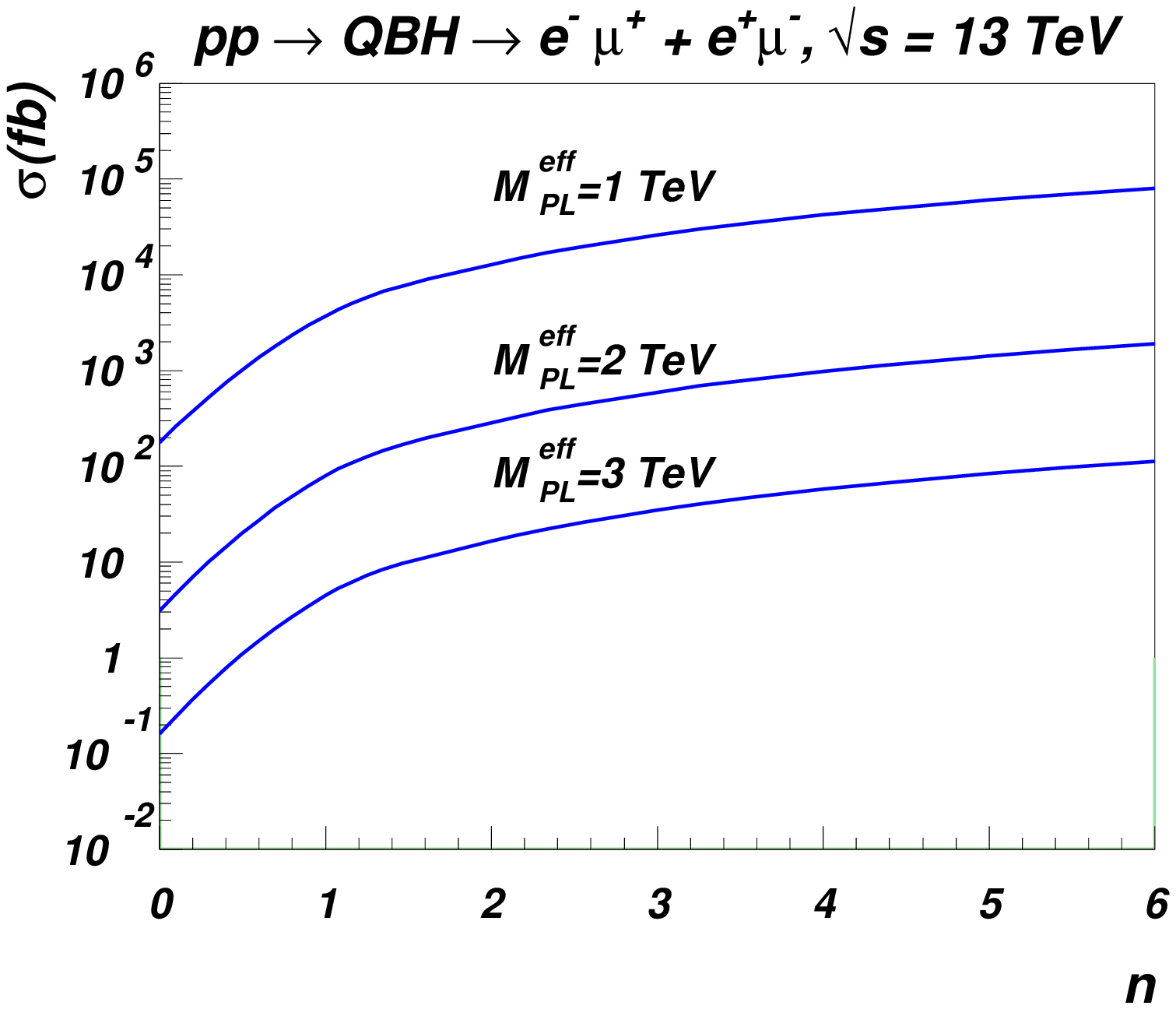}%
\vspace*{-0.5cm}
\caption{\label{fig:qbh-cs}
The cross section of $pp\to QBH(0,0)\to e^-\mu^+ (+e^+\mu^-)$
process at the LHC for 8 TeV (left) and 13 TeV (right) centre-of-mass energy $pp$ collisions.
Upper row: cross section versus  $\overline{M}_{PL}$, bottom row:
cross section versus $n$.} %
\label{cs1}
\end{figure}

\begin{table}
\begin{tabular}{|c|c|l|l|l|l|l|l|l|}
\hline
&$\overline{M}_{PL}$ (TeV)& n=0 &n=1& n=2 & n=3& n=4& n=5&  n=6\\
\cline{2-9}
\hline
\multirow{3}{*}{LHC@8TeV} &
 1.0 & $32.3$		&  $782.$	&  $2760.$	& $5730.$ &$9370.$  	&  $13500.$ & $18100.$	     \\  \cline{2-9}
&2.0 & $0.235$		&  $6.60$	&  $24.5$	& $51.8$ &$85.7 $  	&  $124.$ & $166.$	     \\ \cline{2-9}
&3.0 & $0.00388$	&  $0.116$	&  $0.439$	& $0.939$ &$1.56$  	&  $2.28$ & $3.06$	     \\ \hline\hline
\multirow{3}{*}{LHC@13TeV} &
 1.0  &$177.$	& $373.$& $12800.$& $26000.$& $42200.$ & $60500.$ &$80400.$		 \\  \cline{2-9}
& 2.0 &$3.11$ 	& $79.7$& $286.$& $596.$    & $980.$   & $1420.$ &$1890.$		 \\ \cline{2-9}
& 3.0 &$0.161$  & $4.48$& $16.5$& $34.8$& $57.4$ & $83.5$ &$112.$		  \\ 
\hline\hline\end{tabular}
\caption{\label{tab:cs} The cross section for  $pp\to QBH(0,0)\to e^-\mu^+ (+e^+\mu^-)$
process at the LHC in fb for 8 TeV  and 13 TeV  centre-of-mass energy $pp$ collisions
for $\overline{M}_{PL}$=1,2,3 TeV and $n$=1-6.}
\end{table}

We start by presenting the QBH production cross section in  Fig.~\ref{fig:qbh-cs} where
the cross section versus  $\overline{M}_{PL}$ (upper row) and versus $n$ (bottom row)
is given for $pp\to QBH(0,0)\to e^-\mu^+ (+e^+\mu^-)$
process at the LHC for 8 TeV (left) and 13 TeV (right) centre-of-mass energy $pp$ collisions.
The respective specific numbers for the cross section are given in Table~\ref{tab:cs}.
Note that the cross section for $pp\to QBH(0,0)\to e^-e^+$
production is a factor of  two smaller because of the respective QBH branching ratio.
One can observe a big  difference in cross sections between effective four-dimensional case ($n=0$) and higher dimensional theories, for which the cross section of QBH production
can be  three orders of magnitude higher as, for example, for $n=6$ case,
when the cross section driven by the factor (\ref{eq:ff}). The cross section dependence 
as a function of $n$ is explicitly presented in the bottom 
row of Fig.~\ref{fig:qbh-cs} for three fixed values of $\overline{M}_{PL}=1,2,3$~TeV.
One can note that the steep cross section drop as a function of  $\overline{M}_{PL}$
is defined by PDFs and reaches 0.1 fb around $\overline{M}_{PL}=2$~TeV 
for  $n=0$ and  around $\overline{M}_{PL}=4$~TeV for $n=6$
at the LHC@8TeV. At the LHC@13TeV the cross section reaches 0.1 fb around $\overline{M}_{PL}=3$~TeV 
for  $n=0$ and  around $\overline{M}_{PL}=6$~TeV for $n=6$.
Let us note that 0.1 fb cross section level  is the typical sensitivity 
which is expected at 20 fb$^{-1}$ at 
LHC@8TeV or 30 fb$^{-1}$ at LHC@13TeV (first year run) luminocities providing respectively few
events which under assumption of the negligible background allows to establish exclusion at the 95\% CL.
\begin{figure}[htb]
\begin{center}
\epsfig{width=0.5\textwidth,file=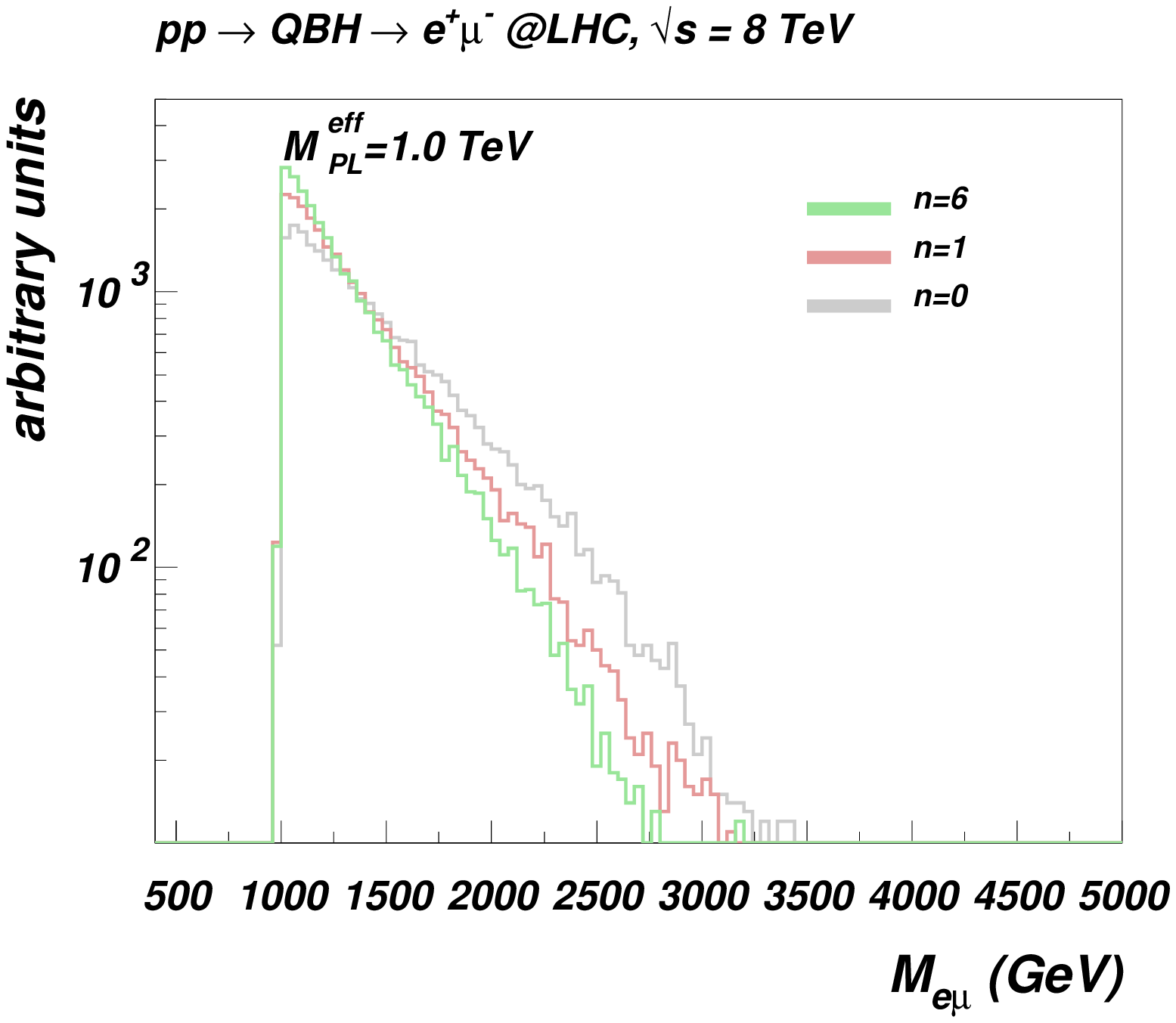}%
\epsfig{width=0.5\textwidth,file=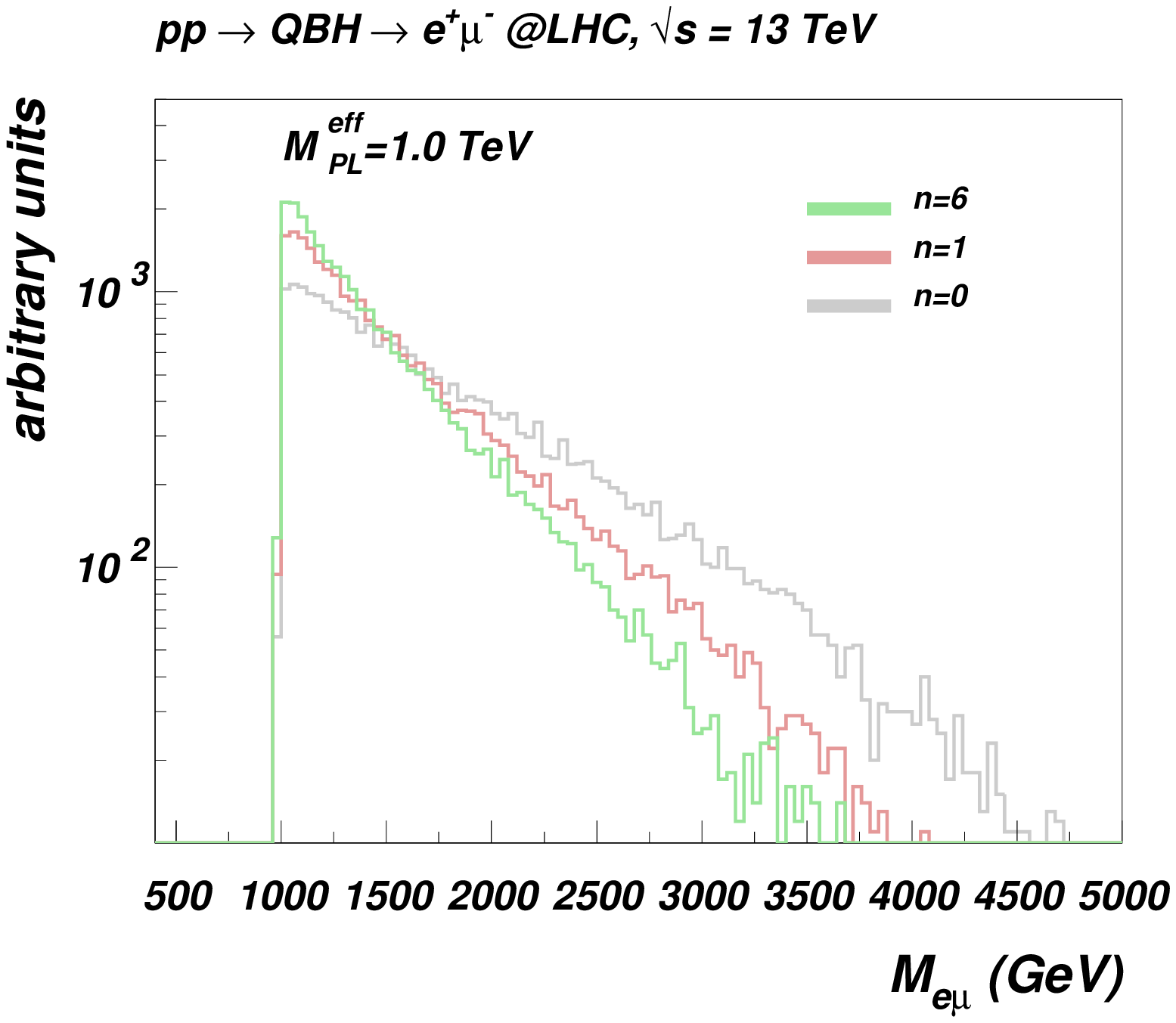}\\
\epsfig{width=0.5\textwidth,file=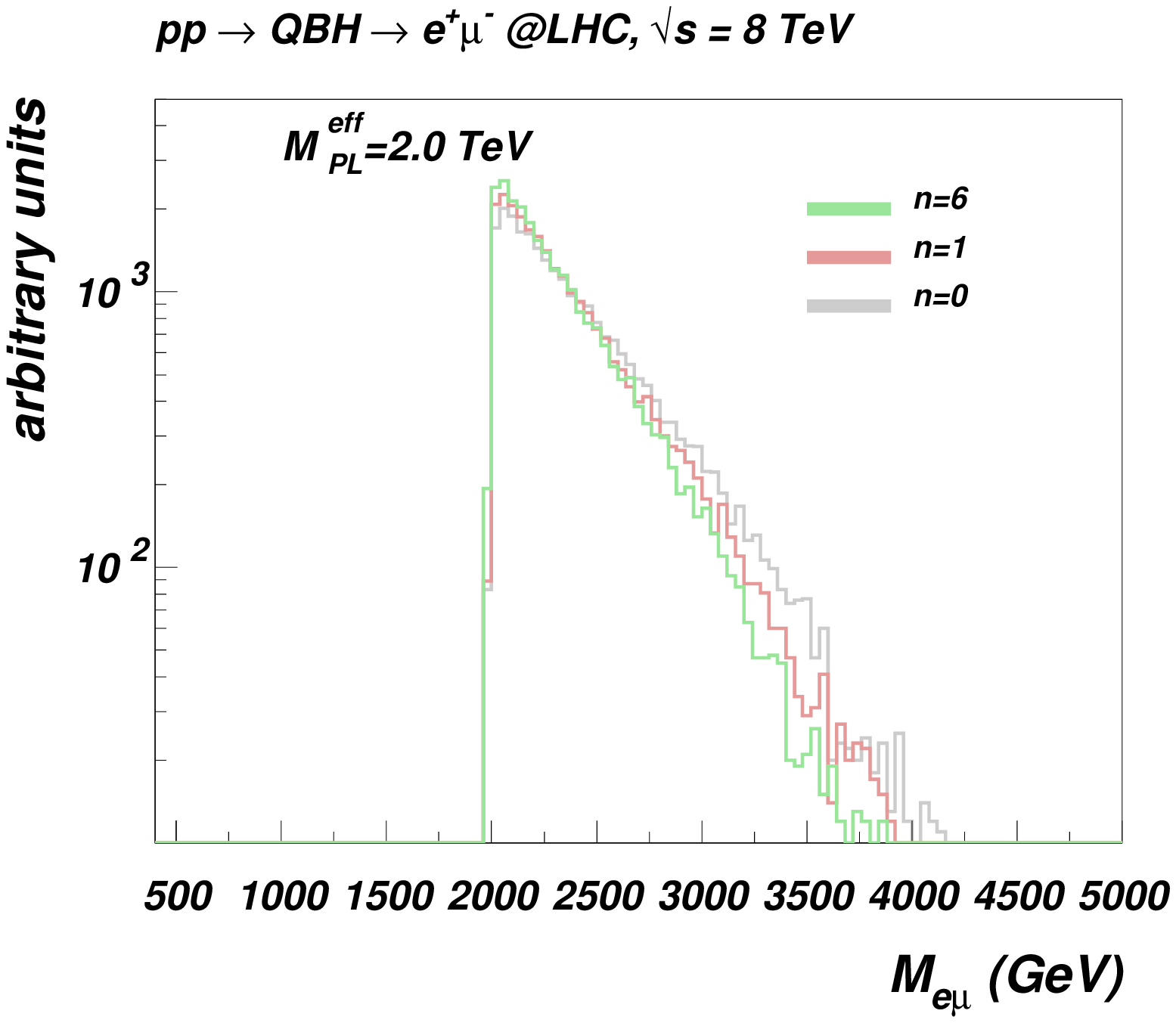}%
\epsfig{width=0.5\textwidth,file=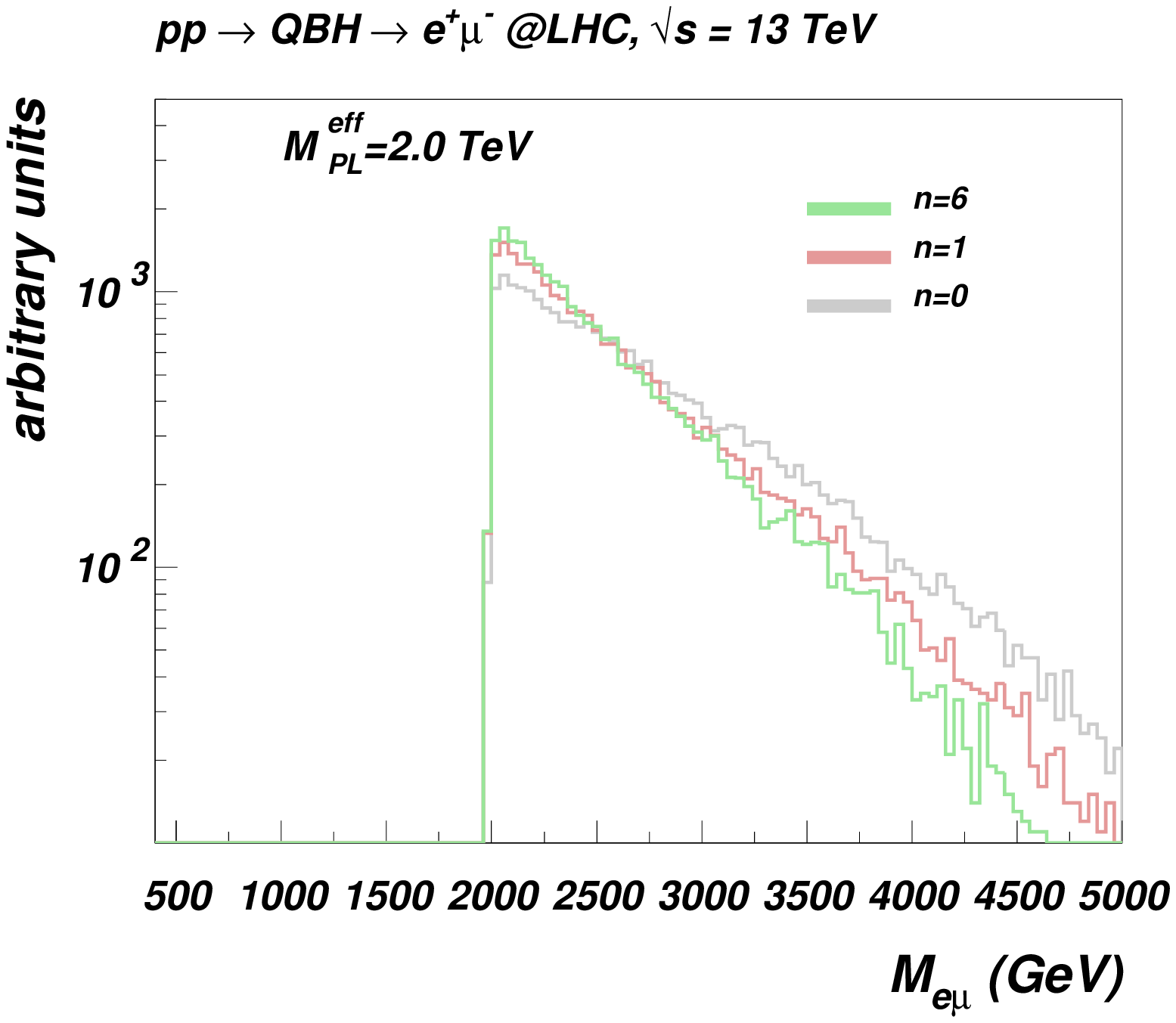}%
\end{center}
\caption{\label{fig:distr1}
Invariant mass of $e\mu$ distribution
for different $n$ for LHC for 8 TeV (left) and 13 TeV (right) 
and  $\overline{M}_{PL}=1$(left) and 2 TeV (right).
Results are presented with the same normalisation} %
\label{cs2}
\end{figure}
It is worth discussing  the {\it shape} of the kinematical distributions from  QBH(0,0) decay products. In Fig.~\ref{fig:distr1} we present  $e\mu$ invariant mass distribution
for different $n$ for LHC for 8 TeV (left) and 13 TeV (right) 
and  $\overline{M}_{PL}=1$(left) and 2 TeV (right).
Results are presented for the same normalisation
to compare the shapes of the distributions for different $n$.
The signal shape exhibits a threshold production nature 
and driven primarily by steeply falling PDFs. It  is qualitatively different from a Breight-Wigner shape of resonances, e.g. $Z'$ bosons, appearing in various BSM models 
different from QBH ones.
One can observe  the shape difference between different extra-dimensional
models and the effective four-dimensional theory. Moreover, the more phase space is available,
the bigger difference in the high invariant mass tail
which drops faster for larger number of extra dimensions.
This is actually  what one can expect recalling the energy dependent nature of the form-factor
given by Eq.(\ref{eq:ff}). So,  $n=0$\footnote{Note that the limit for $n=0$ obtained here does apply to the specific models described in \cite{Calmet:2008tn,Calmet:2010nt,Calmet:2014gya} since these models have large hidden sectors and neutral quantum black holes would decay massively in the particles of the hidden sector. However, charged black holes, which are not considered here, would decay into standard model particles.}
distributions has the slowest $M_{e\mu}$ dependence which sharpens with the  increase of  $n$ driven by Eq.(\ref{eq:ff}).
One can see that in the  large $n$ limit the   parton level asymptotically 
becomes less and less $s$ dependent, so $M_{e\mu}$ distributions become similar
and are defined  by rapidly falling PDFs.
One should also note that 
all $M_{e\mu}$ distributions,  exhibit a clear step at the QBH production threshold
and are qualitatively different from the resonant Breit-Wigner shape.
Therefore in our analysis of the LHC sensitivity to the QBH parameter space, we set a lower 
 $M_{e\mu}$ cut rather than a mass-window cut.
 
\begin{figure}[htb]
\begin{center}
\epsfig{width=0.5\textwidth,file=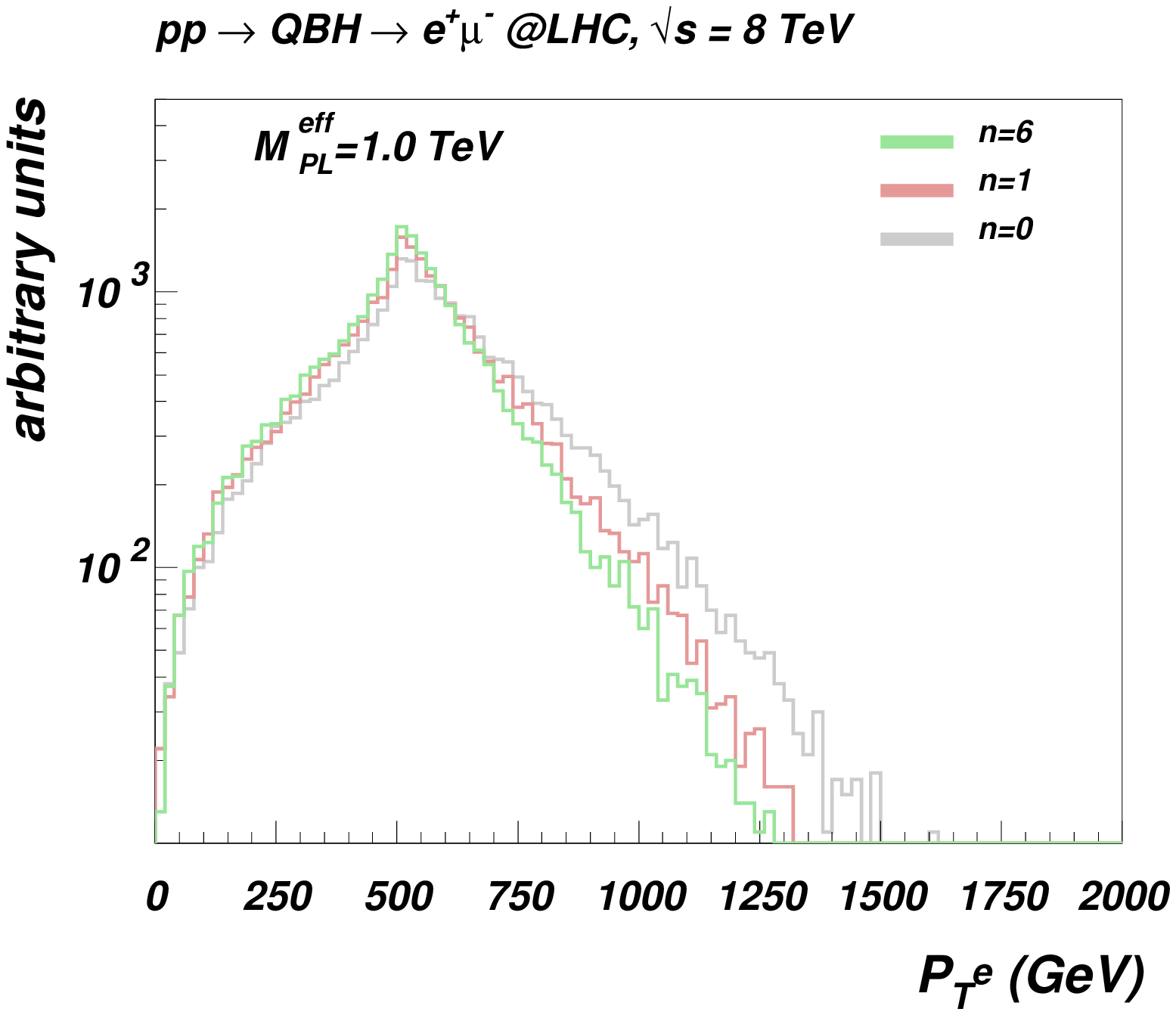}%
\epsfig{width=0.5\textwidth,file=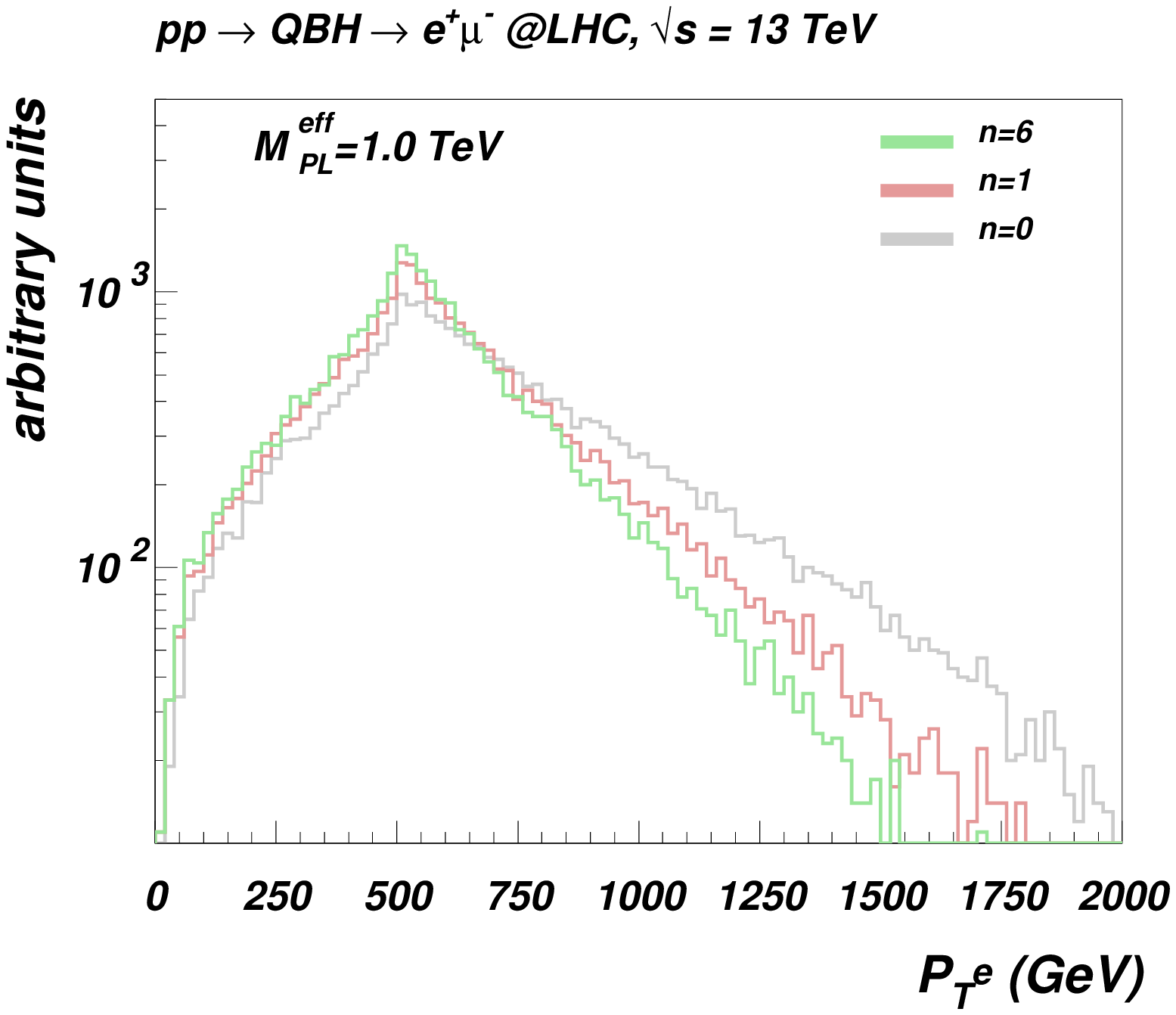}\\
\epsfig{width=0.5\textwidth,file=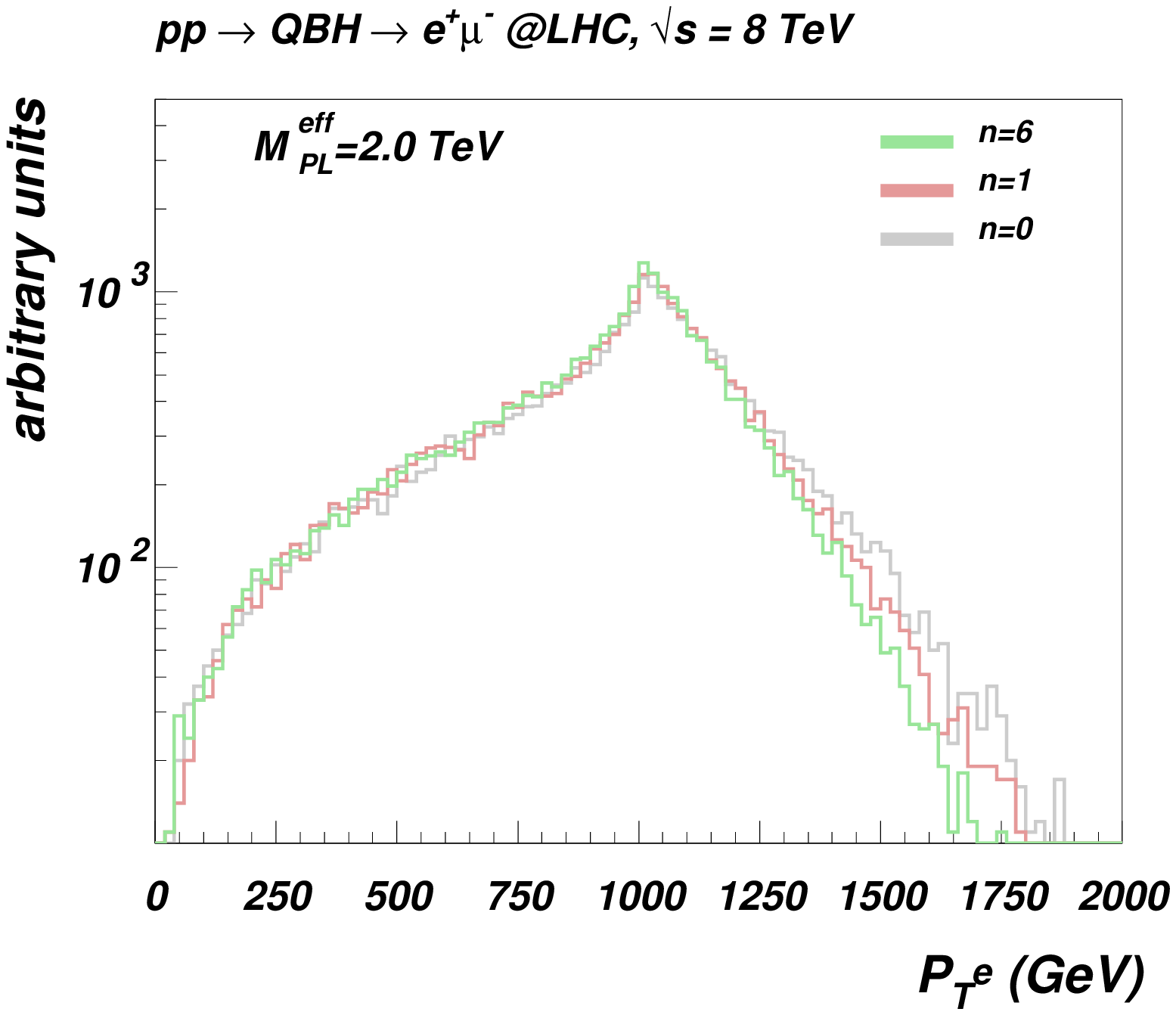}%
\epsfig{width=0.5\textwidth,file=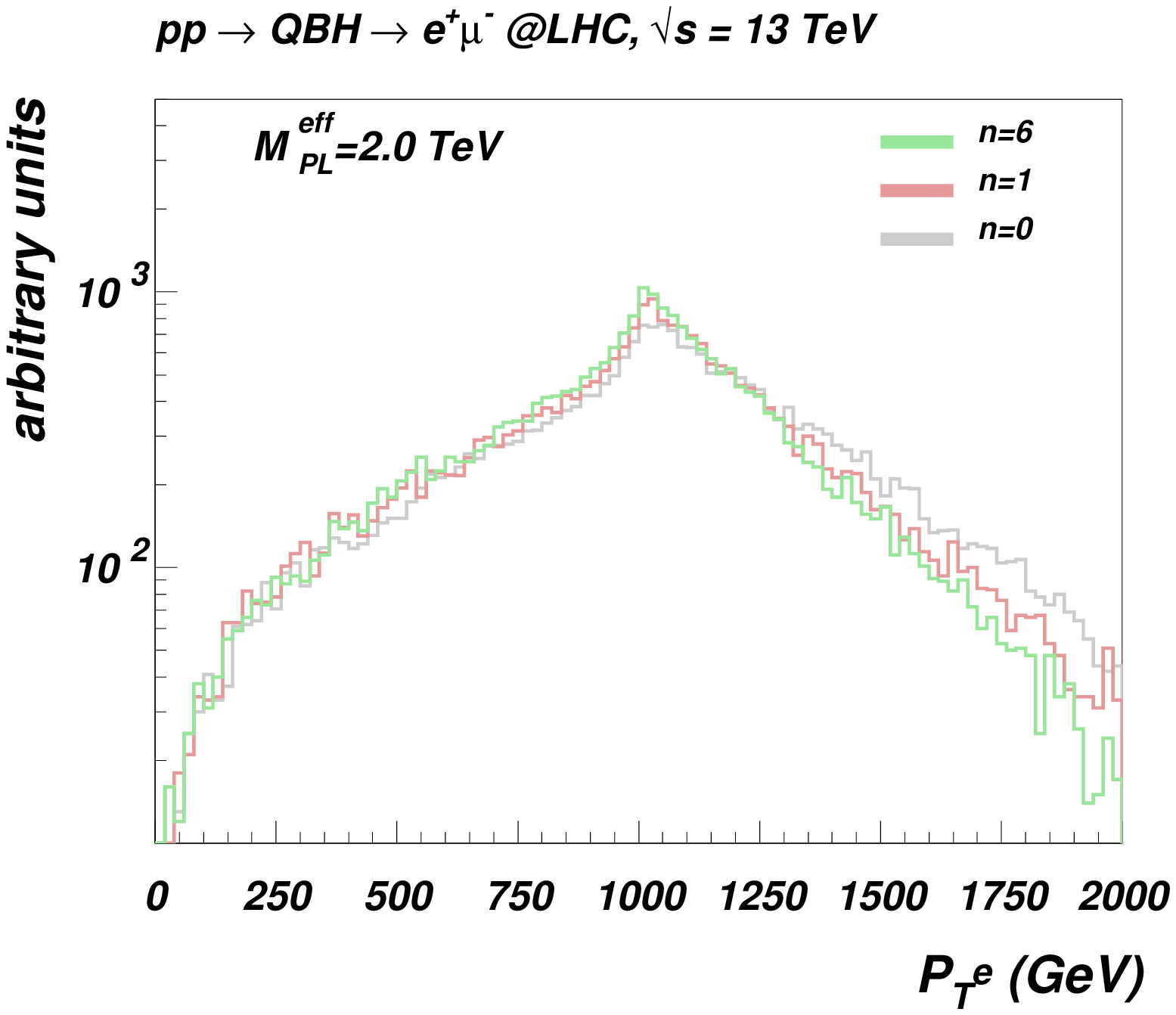}%
\end{center}
\caption{\label{fig:distr2} $p_T^\ell$ distribution for different $n$ for LHC for 8 TeV (left) and 13 TeV (right) 
and  $\overline{M}_{PL}=1$(left) and 2 TeV (right).
Results are presented for the same normalisation} %
\label{cs3}
\end{figure}
\begin{figure}[htb]
\begin{center}
\epsfig{width=0.5\textwidth,file=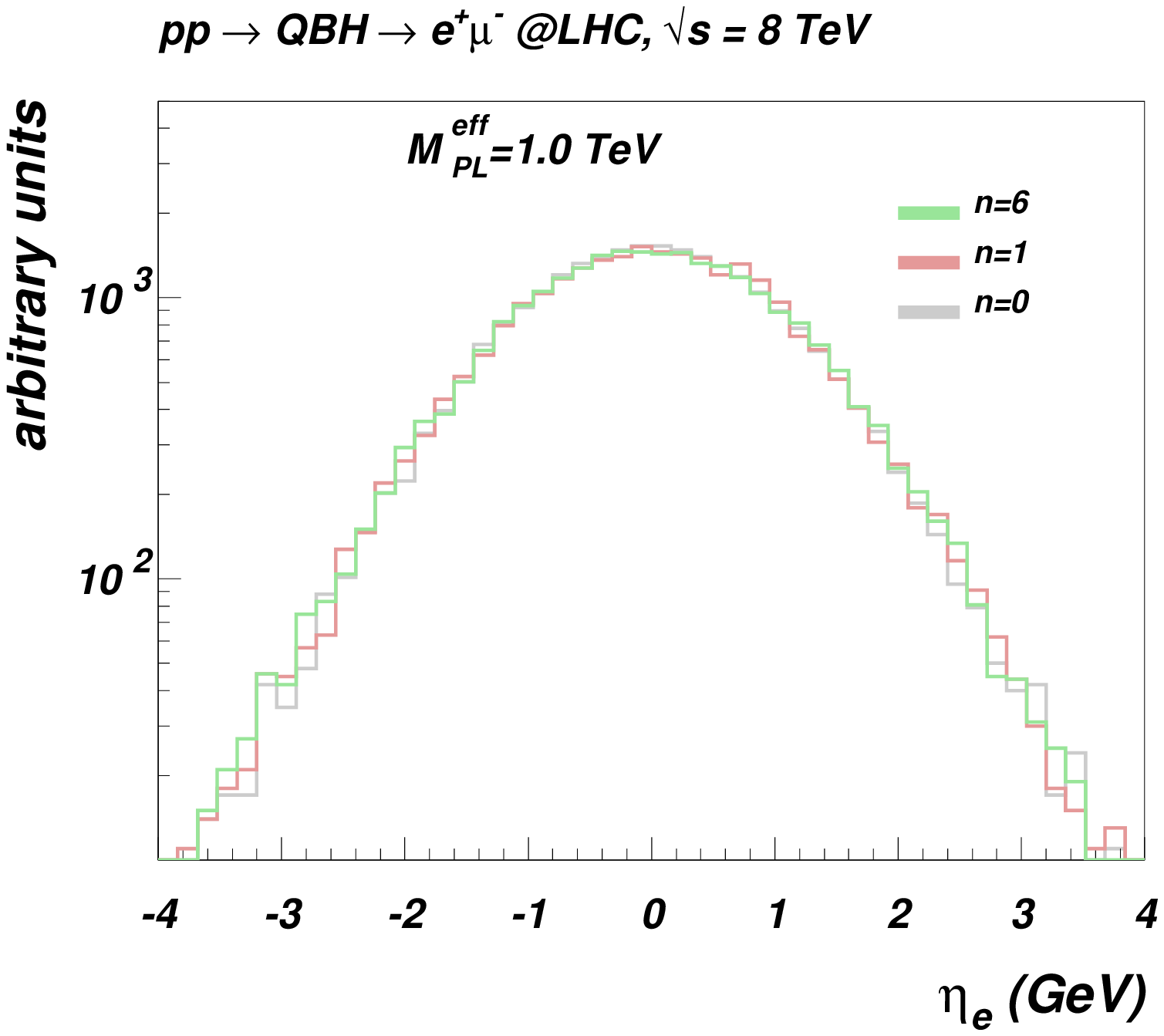}%
\epsfig{width=0.5\textwidth,file=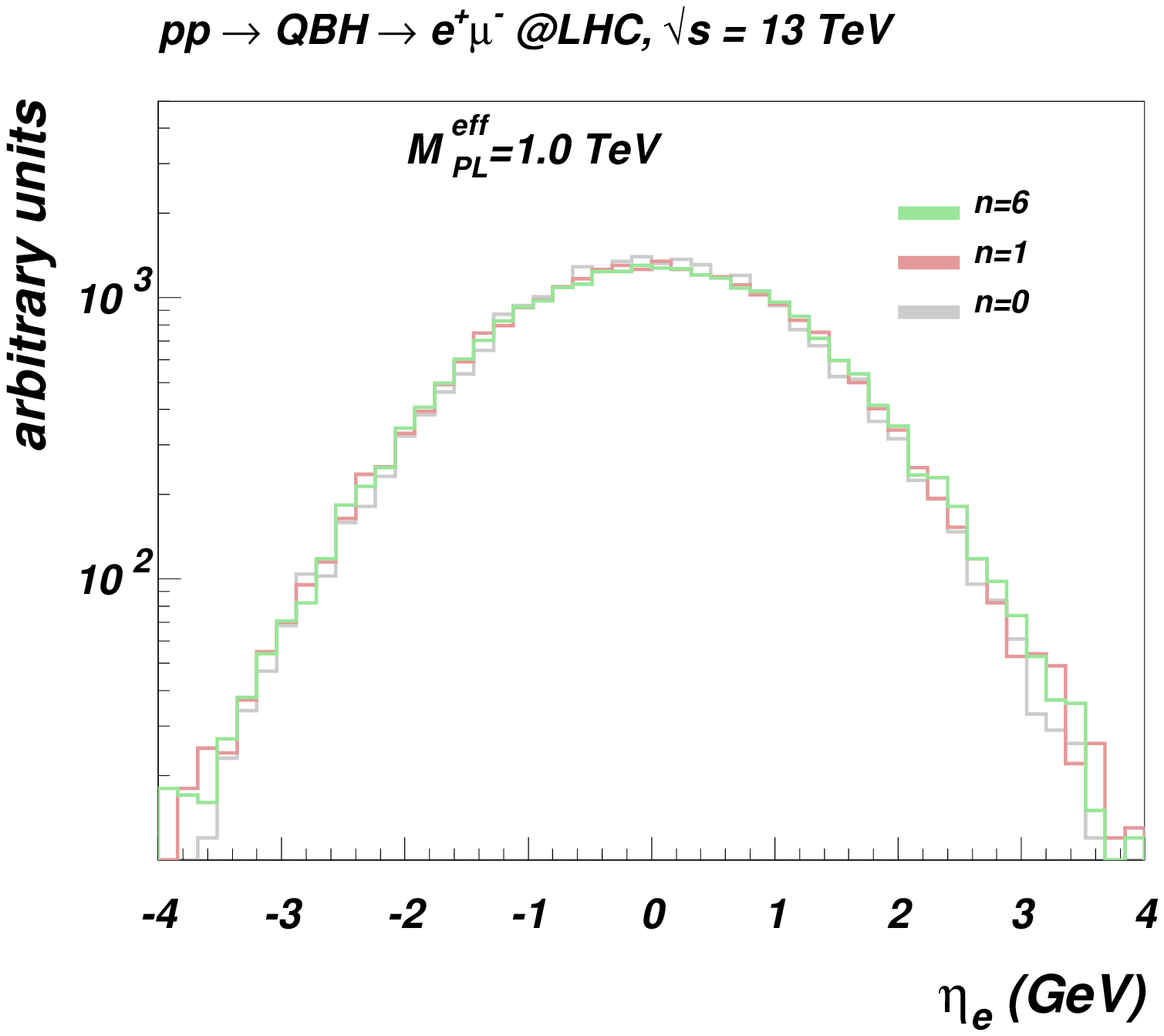}\\
\epsfig{width=0.5\textwidth,file=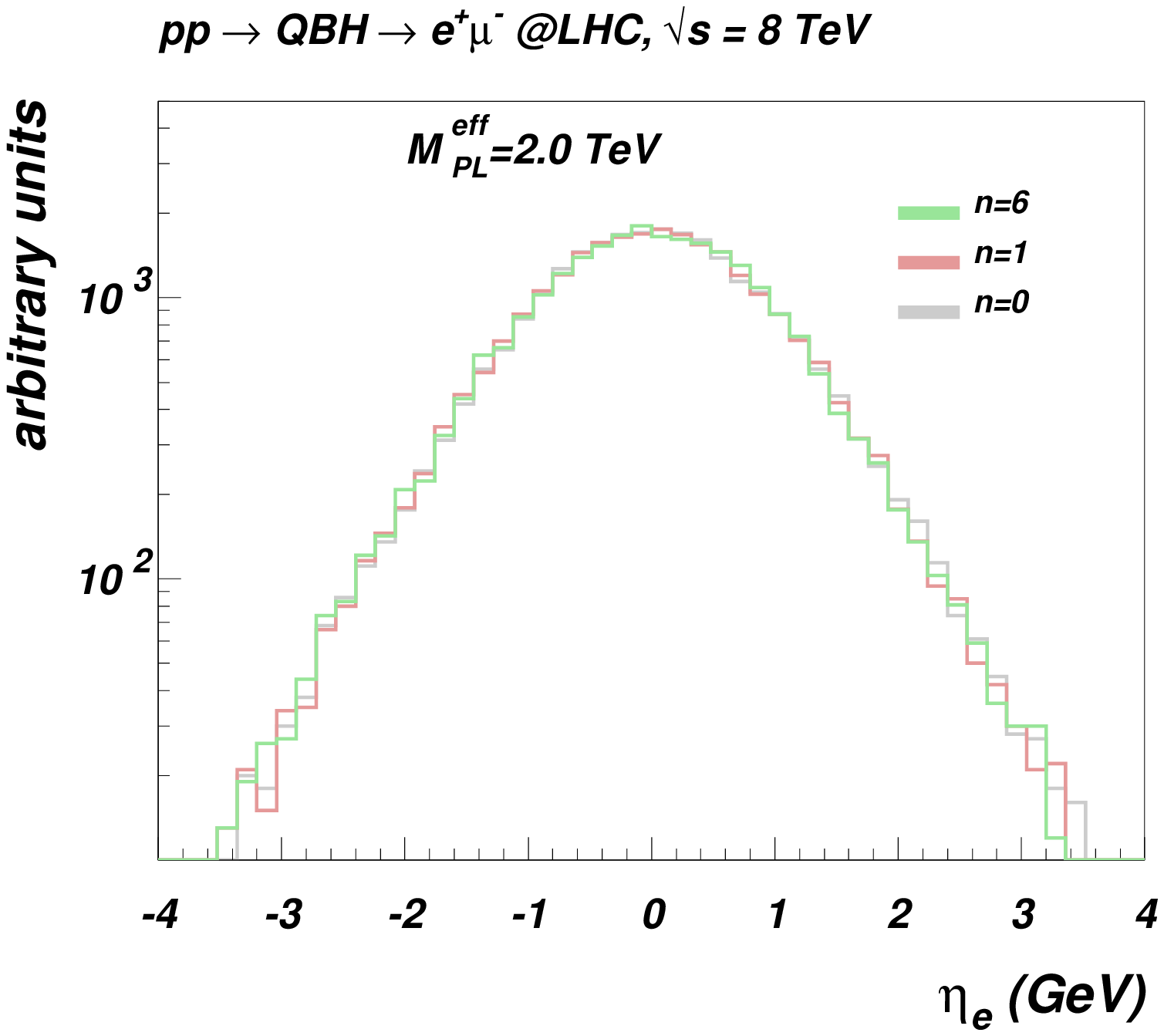}%
\epsfig{width=0.5\textwidth,file=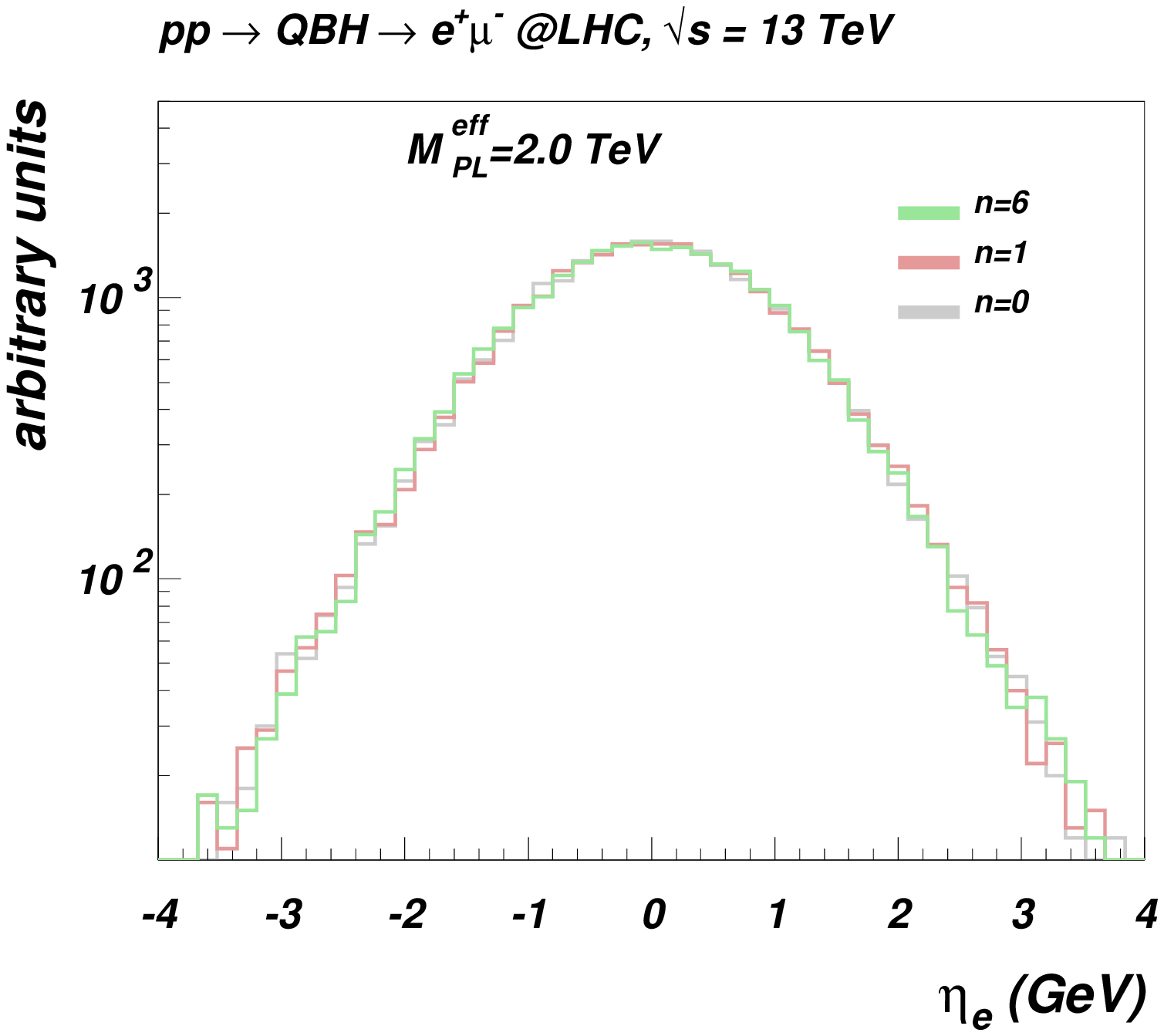}%
\end{center}
\caption{\label{fig:distr3} Pseudorapidity of lepton
 distribution for different $n$ for LHC for 8 TeV (left) and 13 TeV (right) 
and  $\overline{M}_{PL}=1$(left) and 2 TeV (right).
Results are presented for the same normalisation}%
\label{cs}
\end{figure}

Lets turn now to  $p_T^\ell$ distribution presented in  Fig.~\ref{fig:distr2}. One can see that the difference between transverse momentum distributions is clearly connected with $s$ dependence of Eq.(\ref{eq:ff}) and eventually correlated with differences in $M_{e\mu}$ distributions. At the same time it is worth noting that the difference in the high energy 
tail distribution will not visibly affect acceptance/selection cuts as we discuss below.
Finally lets take a look at the pseudorapidity distributions in 
bottom left frame of  Fig.~\ref{fig:distr3} which demonstrate
that $\eta_\ell$ distributions are very similar for the scenarios with  different $n$.
Looking at Fig.\ref{fig:distr1}-\ref{fig:distr3} one can conclude  that  in spite of the differences
for certain  kinematical distributions for different $n$
for high values of $M_{\mu e}$ and $P_T^{e,\mu}$,  one can expect a very similar acceptance efficiency for these models, since all of them provide high $P_T$ leptons
(with $P_T$ far above the acceptance cuts) with a very similar rapidity distributions.

\begin{figure}[htb]
\epsfig{width=0.5\textwidth,file=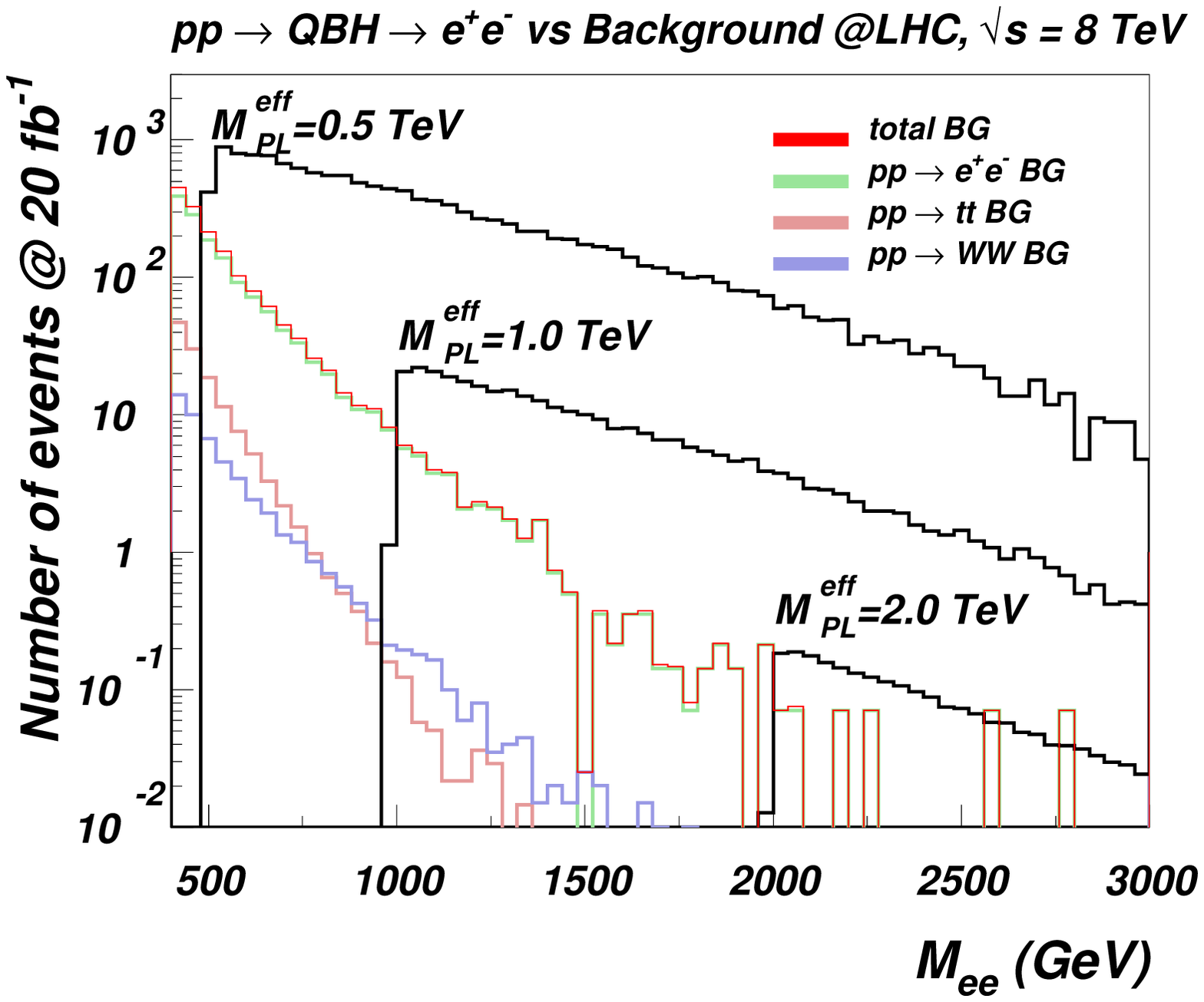}%
\epsfig{width=0.5\textwidth,file=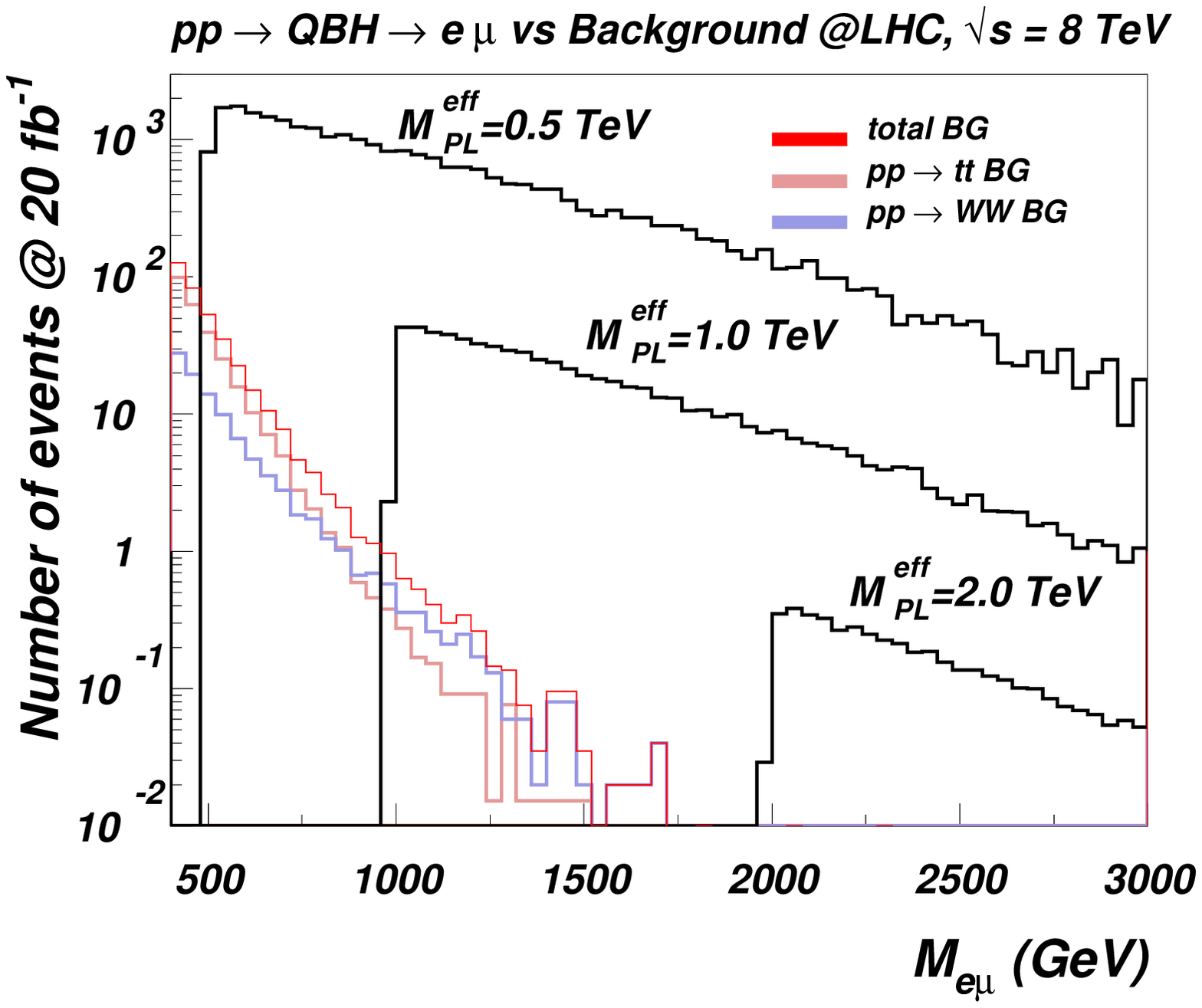}%
\caption{\label{fig:sb8tev} 
 Invariant mass distributions for $e^+e^-$ (left) and $e\mu$ (right) QBH signatures ($n=0$ case)
 and the respective backgrounds for LHC@8TeV}%
\end{figure}

\begin{figure}[htb]
\epsfig{width=0.5\textwidth,file=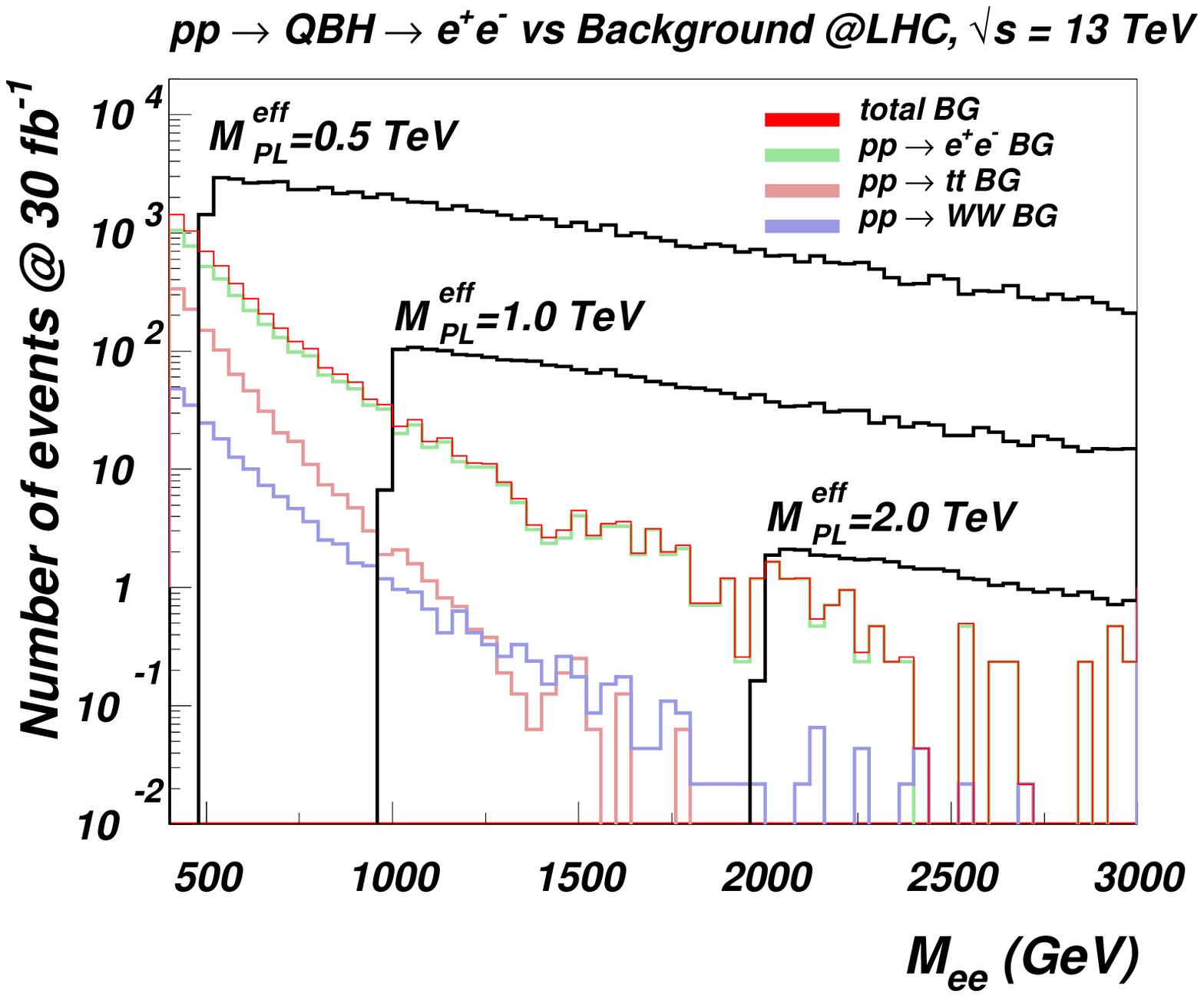}%
\epsfig{width=0.5\textwidth,file=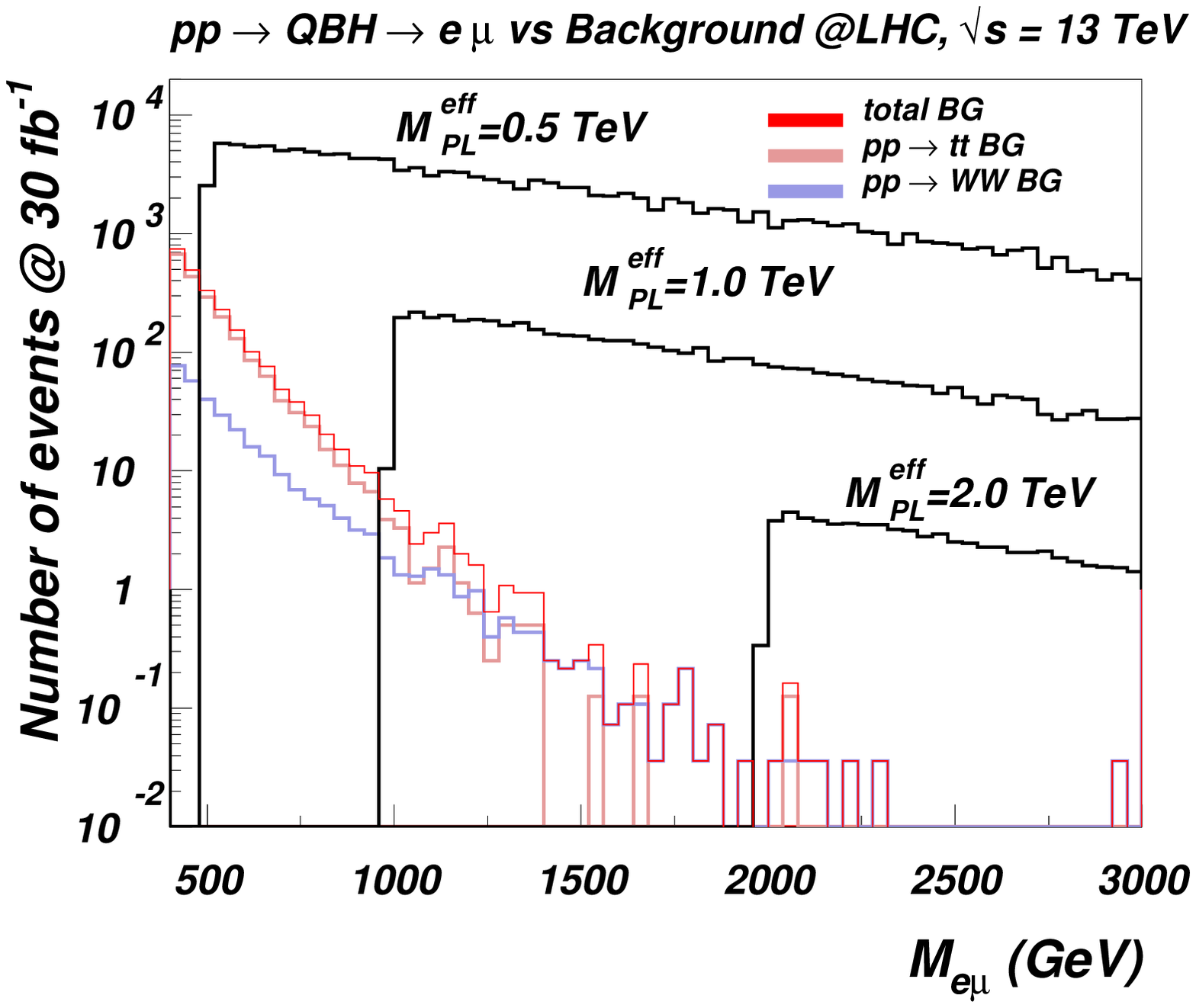}%
\caption{\label{fig:sb13tev} 
 Invariant mass distributions for $e^+e^-$ (left) and $e\mu$ (right) QBH signatures ($n=0$ case) and the respective backgrounds for LHC@13TeV} %
\label{13tev}
\end{figure}

We have also performed signal vs background analysis for the QBH(0,0)
production at the LHC decaying into $e^+e^-$ and $e\mu$ final states.
The main backgrounds for  the $e^+e^-$ signature are $pp \to e^+e^-$ Drell-Yan (DY) process, 
as well as $\bar{t}t$ and  $W^+W^-$ pair production.
The rate of these backgrounds together with the signal rate for  
$\overline{M}_{PL}=0.5$, 1 and 2 TeV is presented in Fig.~\ref{fig:sb8tev}(left)
for $M_{e^+e^-}$ invariant mass distribution. The QBH signal is shown for $n=0$ case.
One can see the dominat DY background is below the signal, but non negligible
for not very high QBH masses.
At the same time   DY background is absent in case  $e\mu$ signature,
as shown in Fig.~\ref{fig:sb8tev}(right).
One can also see that in case of this signature  $\bar{t}t$ and  $W^+W^-$ backgrounds are 
negligible, so LHC potential to probe this signature purely depends on the signal rate
which is defined by $\overline{M}_{PL}$ and $n$ parameters.
Analogous distributions are presented in Fig.~\ref{fig:sb13tev} for LHC@13 TeV
exhibiting qualitatively the same pattern for signal and backgrounds
for the $e^+e^-$ and  $e\mu$ signatures under study.

\begin{figure}[htb]
\centerline{
\epsfig{width=0.8\textwidth,file=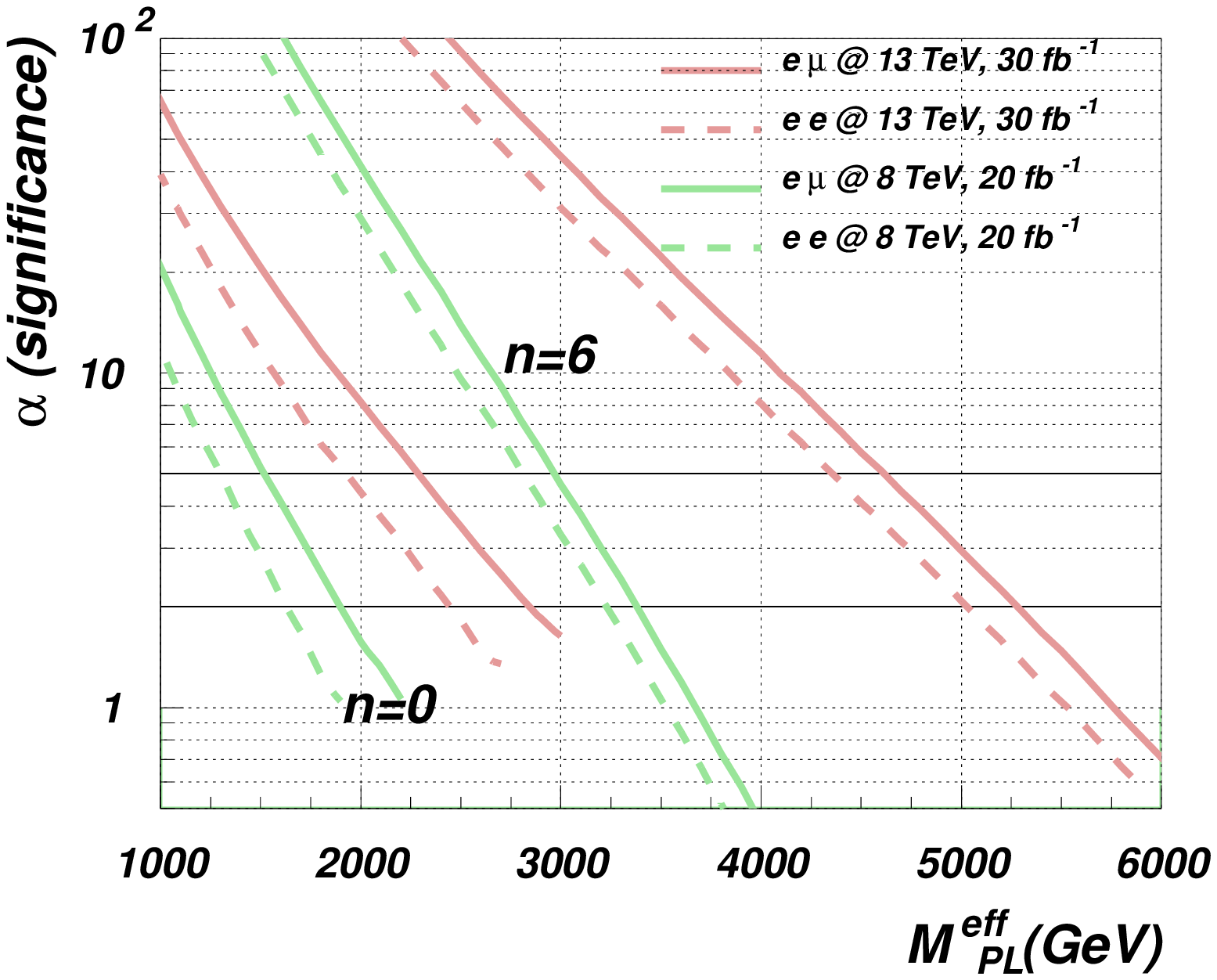}}
\caption{Signal significance
for QBH $e^+e^-$ and $e\mu$ signatures as a function of
 $\overline{M}_{PL}$} for  $n=0$ and 6
 at the LHC@8 and 13 TeV.
\label{fig:alpha}
\end{figure}

At the final step we estimate sensitivity of the LHC@8 and 13 TeV
to the signatures from QBH under study and estimate the respective limits 
in case if signal is not observed.
In our analysis though we restrict ourselves to the study at parton-level and do take into account realistic electromagnetic energy resolution, using a value of
$0.15/\sqrt{E(\text{GeV})}$, which is typical for the ATLAS and CMS detectors
and require $|\eta_{\mu,e}|<2.5$ and $p_T^{e,\mu}$ with respect to the acceptance cuts. %
We also suggest the simple  analysis cut  to be $M_{e\mu}(M_{ee})> 1.1\times \MPL$, noting
that  the acceptance efficiency will be very similar for different $n$ models.
In Fig.~\ref{fig:alpha} we present the signal significance
for QBH signatures under study at the LHC.
For both criteria, exclusion and discovery,
we use the following formula for statistical signal significance $\alpha$ as~\cite{Bityukov:1998zn}
\begin{eqnarray}
\alpha=2(\sqrt{N_S+N_B}-\sqrt{N_B})
\label{eq:signif}
\end{eqnarray}
and require $\alpha\ge 2$ for exclusion region and $\alpha\ge 5$
for the discovery region. The $N_{S(B)}= \sigma_{S(B)} \mathcal{L}$ 
denotes the number of signal (background) events for an integrated luminosity
$\mathcal{L}$.
The figure presents results for  LHC@8 (20 fb$^{-1}$) and 13 TeV (30 fb$^{-1}$) for  $n=0$ and 6 as two extreme cases for the range of $n$ under study.
The 
\begin{table}
\begin{tabular}{|l | l|c|c||c|c|}
\hline
\multicolumn{2}{|c}{} & \multicolumn{2}{|c||}{LHC@8TeV} & \multicolumn{2}{|c|}{LHC@13TeV} \\
\hline
CL & n &$e^+e^-$ & $e\mu$ &  $e^+e^-$ & $e\mu$\\
\hline
\hline
95\%CL &0	&1920 GeV &2240 GeV &3360 GeV&3780 GeV 
\\ 
\hline
5$\sigma$&0	&1550 GeV &1810 GeV &2790 GeV &3180 GeV 
\\  
\hline
95\%CL 	&6	&3540 GeV &3680 GeV &5510 GeV &5750 GeV  
\\  
\hline
5$\sigma$&6	&3140 GeV &3300 GeV &4850 GeV &5100 GeV 
\\ 
\hline
\hline
\end{tabular}
\caption{\label{tab:limits}
$\overline{M}_{PL}$ exclusion and discovery limits for  LHC@8 and 13 TeV for $n=0$ and $n=6$
scenarios.}
\end{table}

The respective $\overline{M}_{PL}$ exclusion and discovery limits for  LHC@8 and 13 TeV for $n=0$ and $n=6$
scenarios are presented in Table~\ref{tab:limits}.
One can see that for $n=0$ the LHC@8TeV the limit on  $\overline{M}_{PL}$ @95\% CL 
is only about 1.92 TeV for $e^+e^-$ signature and 2.24 TeV for  $e\mu$  one,
while the respective discovery numbers are 1.55 and 1.81 TeV respectively.
The  LHC@13TeV in the first year will be able to improve limits and cover
$\overline{M}_{PL}$ @95\% CL up to 3.36 TeV with  $e^+e^-$ signature 
and up to 3.78 TeV with  $e\mu$, and discover QBH with $\overline{M}_{PL}$
up to 2.79 and 3.18 TeV respectively.
At the same time the for $n=6$  for which the QBH  production cross section is about 3 orders of magnitude higher, the LHC reach is much more impressive.
For example, with  $e\mu$ signature  LHC@8 will be able to exclude 
$\overline{M}_{PL}<3.68$~TeV @95\% CL or discover QBH with $\overline{M}_{PL}<3.30$~TeV.
Analogous numbers for  LHC@13TeV are even more exciting -- it would be able to probe 
$\overline{M}_{PL}<5.75$~TeV or discover the $e\mu$  signal for  $\overline{M}_{PL}<5.10$~TeV.
In is worth noting that though our analysis reproduces quite well recent ATLAS results
on QBH search  at LHC@8TeV~\cite{Aad:2014cka}, which stated the $3.65$~TeV limit for $n=6$ case
for $e^+e^- + \mu^+\mu^-$ signatures. 
Since the signal cross section for this signature
equal to the cross section for the  $e\mu$ signal while background is negligible,
the limits are expected to be the same for both cases.
The respective limit from our study is $3.67$~TeV which is in a very good agreement 
with the above on from ATLAS.
We should also mention that the signal  cross section, quoted by \cite{Aad:2014cka} for  $n=6$ case
agrees  within 10\% with the cross section we found in our paper.
Therefore we can also conclude  about  the successful validation of our generator and analysis
for the LHC@8TeV.

%%%%%%%%%%%%%%%%%%%%
%%%%%%%%%%%%%%%%%%%%
%%%%%%%%%%%%%%%%%%%%
%%%%%%%%%%%%%%%%%%%%

\section{Conclusions}

We discuss a field theoretical framework to describe the interactions of non-thermal QBH with particles of the Standard Model and  propose a non-local Lagrangian to describe the production of these QBH  which is designed  to reproduce the geometrical cross section $\pi r_s^2$ for black hole production.
We have implemented this model  into CalcHEP and it is  publicly available at the High Energy Model Database for simulation of QBH events at the LHC and future colliders. This model, QBH@HEPMDB is an effective independent tool for QBH phenomenological and experimental explorations.
Detailed comparison of QBH@HEPMDB with analogous  tools on the market requires  dedicated work which we plan to perform 
in the nearest future.
In this paper we present the first phenomenological application of the QBH@HEPMDB
model with spin-0 neutral QBH giving rise the  $e^+e^-$ and $e\mu$  signatures at the LHC@8TeV and LHC@13TeV and
produce the first respective projections
in terms of limits on the reduced Planck mass, $\overline{M}_{PL}$ and the number of the
extra-dimensions $n$. In particular we found that among two signatures, $e\mu$ one provides the best LHC reach since it is free of DY
background. 
We have successfully validated our generator and exclusion limits against recent ATLAS results for $n=6$ case.
We found that with $e\mu$   signature, for number of extra-dimensions, $n$, in the range of 0-6, the
LHC@8 will be able to probe the respective range of 2.2-3.7 TeV of the reduced Planck Mass $\overline{M}_{PL}$.
We have also produced new projections for  LHC@13 and  found that even in the first year of operation
with 30$fb^{-1}$ the range  3.8-5.8 TeV of  $\overline{M}_{PL}$
at 95\%CL can be probed.
The respective discovery range of  LHC@13 is 3.2-5.1 TeV for $\overline{M}_{PL}$.
\\
\\
{\it Acknowledgements:}
We are grateful  to Tim Morris, Alexander Pukhov, Claire Shepherd-Themistocleous, Emmanuel Olaiya, Andreas Gueth and Thomas Reis for stimulating  discussions.
This work is supported in part by the European Cooperation in Science and Technology (COST) action MP0905 ``Black Holes in a Violent  Universe" and by the Science and Technology Facilities Council, grants number  ST/L000504/1  and ST/L000296/1.
%%%%%%%%%%%%%%%%%%%%
%%%%%%%%%%%%%%%%%%%%
%%%%%%%%%%%%%%%%%%%%
%%%%%%%%%%%%%%%%%%%%

%%%%%%%%%%%%%%%%%%%%%%%%%%%%%%%%%%%%%%%%%%%%%%%%%%%%%%%%%%%%%%%%%
%%%
%%%                     BIBLIOGRAPHY
%%%
%%%%%%%%%%%%%%%%%%%%%%%%%%%%%%%%%%%%%%%%%%%%%%%%%%%%%%%%%%%%%%%%%


\begin{thebibliography}{0}
\expandafter\ifx\csname natexlab\endcsname\relax\def\natexlab#1{#1}\fi
\expandafter\ifx\csname bibnamefont\endcsname\relax
  \def\bibnamefont#1{#1}\fi
\expandafter\ifx\csname bibfnamefont\endcsname\relax
  \def\bibfnamefont#1{#1}\fi
\expandafter\ifx\csname citenamefont\endcsname\relax
  \def\citenamefont#1{#1}\fi
\expandafter\ifx\csname url\endcsname\relax
  \def\url#1{\texttt{#1}}\fi
\expandafter\ifx\csname urlprefix\endcsname\relax\def\urlprefix{URL }\fi
\providecommand{\bibinfo}[2]{#2}
\providecommand{\eprint}[2][]{\url{#2}}

\end{thebibliography}


\bigskip

\baselineskip=1.6pt
\begin{thebibliography}{99}

%\cite{ArkaniHamed:1998rs}
\bibitem{ArkaniHamed:1998rs}
  N.~Arkani-Hamed, S.~Dimopoulos, G.~R.~Dvali,
  %``The Hierarchy problem and new dimensions at a millimeter,''
  Phys.\ Lett.\  {\bf B429}, 263-272 (1998)
  [hep-ph/9803315];
  %\cite{Antoniadis:1998ig}
%\bibitem{Antoniadis:1998ig}
  I.~Antoniadis, N.~Arkani-Hamed, S.~Dimopoulos {\it et al.},
  %``New dimensions at a millimeter to a Fermi and superstrings at a TeV,''
  Phys.\ Lett.\  {\bf B436}, 257-263 (1998)
  [hep-ph/9804398].
  
  %\cite{Randall:1999ee}
\bibitem{Randall:1999ee}
  L.~Randall, R.~Sundrum,
  %``A Large mass hierarchy from a small extra dimension,''
  Phys.\ Rev.\ Lett.\  {\bf 83}, 3370-3373 (1999)
  [hep-ph/9905221].
  
  %\cite{Calmet:2008tn,Calmet:2010nt,Calmet:2014gya}
  
  %\cite{Calmet:2008tn}
\bibitem{Calmet:2008tn}
  X.~Calmet, S.~D.~H.~Hsu, D.~Reeb,
  %``Quantum gravity at a TeV and the renormalization of Newton's constant,''
  Phys.\ Rev.\  {\bf D77}, 125015 (2008)
  [arXiv:0803.1836 [hep-th]].
  
  
  %\cite{Calmet:2010nt}
\bibitem{Calmet:2010nt}
  X.~Calmet,
  %``A review of Quantum Gravity at the Large Hadron Collider,''
  Mod.\ Phys.\ Lett.\  A {\bf 25}, 1553 (2010)
  [arXiv:1005.1805 [hep-ph]].
  %%CITATION = MPLAE,A25,1553;%%
  
  
    %\cite{Calmet:2014gya}
\bibitem{Calmet:2014gya} 
  X.~Calmet,
  %``The Lightest of Black Holes,''
  arXiv:1410.2807 [hep-th].
  %%CITATION = ARXIV:1410.2807;%%
  

 
  
  
    %\cite{Dimopoulos:2001hw,Giddings:2001bu}
    %\cite{Dimopoulos:2001hw}
\bibitem{Dimopoulos:2001hw}
  S.~Dimopoulos and G.~L.~Landsberg,
  %``black holes at the LHC,''
  Phys.\ Rev.\ Lett.\  {\bf 87}, 161602 (2001)
  [arXiv:hep-ph/0106295].
  %%CITATION = PRLTA,87,161602;%%
 
  %\cite{Banks:1999gd}
\bibitem{Banks:1999gd}
  T.~Banks and W.~Fischler,
  %``A model for high energy scattering in quantum gravity,''
  arXiv:hep-th/9906038.
  %%CITATION = HEP-TH/9906038;%%
  
  %\cite{Giddings:2001bu}
\bibitem{Giddings:2001bu}
  S.~B.~Giddings and S.~D.~Thomas,
  %``High energy colliders as black hole factories: The end of short  distance
  %physics,''
  Phys.\ Rev.\  D {\bf 65}, 056010 (2002)
  [arXiv:hep-ph/0106219].
  %%CITATION = PHRVA,D65,056010;%%

 %\cite{Dai:2007ki}
\bibitem{Dai:2007ki} 
  D.~-C.~Dai, G.~Starkman, D.~Stojkovic, C.~Issever, E.~Rizvi and J.~Tseng,
  %``BlackMax: A black-hole event generator with rotation, recoil, split branes, and brane tension,''
  Phys.\ Rev.\ D {\bf 77}, 076007 (2008)
  [arXiv:0711.3012 [hep-ph]].
  %%CITATION = ARXIV:0711.3012;%%

   %\cite{Calmet:2008dg}
\bibitem{Calmet:2008dg}
  X.~Calmet, W.~Gong, S.~D.~H.~Hsu,
  %``Colorful quantum black holes at the LHC,''
  Phys.\ Lett.\  {\bf B668}, 20-23 (2008)
  [arXiv:0806.4605 [hep-ph]].
  

%\cite{Feng:2001ib}
\bibitem{Feng:2001ib}
  J.~L.~Feng and A.~D.~Shapere,
  %``Black hole production by cosmic rays,''
  Phys.\ Rev.\ Lett.\  {\bf 88}, 021303 (2002)
  [arXiv:hep-ph/0109106].
  %%CITATION = PRLTA,88,021303;%%

   %\cite{Anchordoqui:2003ug}
\bibitem{Anchordoqui:2003ug}
  L.~A.~Anchordoqui, J.~L.~Feng, H.~Goldberg and A.~D.~Shapere,
  %``Inelastic black hole production and large extra dimensions,''
  Phys.\ Lett.\  B {\bf 594}, 363 (2004)
  [arXiv:hep-ph/0311365].
  %%CITATION = PHLTA,B594,363;%%

 %\cite{Anchordoqui:2001cg}
\bibitem{Anchordoqui:2001cg}
  L.~A.~Anchordoqui, J.~L.~Feng, H.~Goldberg and A.~D.~Shapere,
  %``Black holes from cosmic rays: Probes of extra dimensions and new limits  on
  %TeV-scale gravity,''
  Phys.\ Rev.\  D {\bf 65} (2002) 124027
  [arXiv:hep-ph/0112247].
  %%CITATION = PHRVA,D65,124027;%%
 
   %\cite{Anchordoqui:2003jr}
\bibitem{Anchordoqui:2003jr}
  L.~A.~Anchordoqui, J.~L.~Feng, H.~Goldberg and A.~D.~Shapere,
  %``Updated limits on TeV-scale gravity from absence of neutrino cosmic ray
  %showers mediated by black holes,''
  Phys.\ Rev.\  D {\bf 68}, 104025 (2003)
  [arXiv:hep-ph/0307228].
  %%CITATION = PHRVA,D68,104025;%%
  
  
    %\cite{Calmet:2008rv}
\bibitem{Calmet:2008rv} 
  X.~Calmet and M.~Feliciangeli,
  %``Bound on four-dimensional Planck mass,''
  Phys.\ Rev.\ D {\bf 78}, 067702 (2008)
  [arXiv:0806.4304 [hep-ph]].
  %%CITATION = ARXIV:0806.4304;%%
  %12 citations counted in INSPIRE as of 31 Oct 2014
  %\cite{Calmet:2012mf}
\bibitem{Calmet:2012mf} 
  X.~Calmet, L.~I.~Caramete and O.~Micu,
  %``Quantum Black Holes from Cosmic Rays,''
  JHEP {\bf 1211}, 104 (2012)
  [arXiv:1204.2520 [hep-ph]].
  %%CITATION = ARXIV:1204.2520;%%
  %5 citations counted in INSPIRE as of 31 Oct 2014
%\cite{Arsene:2013nca}
\bibitem{Arsene:2013nca} 
  N.~Arsene, X.~Calmet, L.~I.~Caramete and O.~Micu,
  %``Back-to-Back Black Holes decay Signature at Neutrino Observatories,''
  Astropart.\ Phys.\  {\bf 54}, 132 (2014)
  [arXiv:1303.4603 [hep-ph]].
  %%CITATION = ARXIV:1303.4603;%%
  %3 citations counted in INSPIRE as of 31 Oct 2014
  
  \bibitem{calmetbook1}
X.~Calmet, B.~Carr and E.~Winstanley, ``Quantum Black Holes, '' Springer Briefs in Physics, ISBN-10: 3642389384, ISBN-13: 978-3642389382, Spinger, Edition: 2013.
 


   %\cite{Meade:2007sz}
\bibitem{Meade:2007sz}
  P.~Meade and L.~Randall,
  %``Black Holes and Quantum Gravity at the LHC,''
  JHEP {\bf 0805}, 003 (2008)
  [arXiv:0708.3017 [hep-ph]].
  %%CITATION = JHEPA,0805,003;%%



  

    % \cite{Eardley:2002re,D'Eath:1992hb,Hsu:2002bd,Dimopoulos:2001hw,Banks:1999gd,Giddings:2001bu,Feng:2001ib,Anchordoqui:2003ug,Anchordoqui:2001cg,Anchordoqui:2003jr,Dai:2007ki}
    
  %\cite{Eardley:2002re}
  \bibitem{Eardley:2002re}
  D.~M.~Eardley and S.~B.~Giddings,
  %``Classical black hole production in high-energy collisions,''
  Phys.\ Rev.\  D {\bf 66}, 044011 (2002)
  [arXiv:gr-qc/0201034];
  %%CITATION = PHRVA,D66,044011;%%


   %\cite{D'Eath:1992hb}
  \bibitem{D'Eath:1992hb}
  P.~D.~D'Eath and P.~N.~Payne,
  %``Gravitational Radiation In High Speed Black Hole Collisions. 1.
  %Perturbation Treatment Of The Axisymmetric Speed Of Light Collision,''
  Phys.\ Rev.\  D {\bf 46}, 658 (1992);
    %%CITATION = PHRVA,D46,658;%%
  %\cite{D'Eath:1992hd}
  %\bibitem{D'Eath:1992hd}
  %P.~D.~D'Eath and P.~N.~Payne,
  %``Gravitational Radiation In High Speed Black Hole Collisions. 2. Reduction
  %To Two Independent Variables And Calculation Of The Second Order News
  %Function,''
  Phys.\ Rev.\  D {\bf 46}, 675 (1992);
  %%CITATION = PHRVA,D46,675;%%
  %\cite{D'Eath:1992qu}
  %\bibitem{D'Eath:1992qu}
  %  P.~D.~D'Eath and P.~N.~Payne,
  %``Gravitational radiation in high speed black hole collisions. 3. Results and
  %conclusions,''
  Phys.\ Rev.\  D {\bf 46}, 694 (1992).
  %%CITATION = PHRVA,D46,694;%%
  
  
  %\cite{Hsu:2002bd}
  \bibitem{Hsu:2002bd}
  S.~D.~H.~Hsu,
  %``Quantum production of black holes,''
  Phys.\ Lett.\  B {\bf 555}, 92 (2003)
  [arXiv:hep-ph/0203154].
  %%CITATION = PHLTA,B555,92;%%
  
%\cite{Calmet:2011ta}
\bibitem{Calmet:2011ta} 
  X.~Calmet, D.~Fragkakis and N.~Gausmann,
  %``The flavor of quantum gravity,''
  Eur.\ Phys.\ J.\ C {\bf 71}, 1781 (2011)
  [arXiv:1105.1779 [hep-ph]];
  %%CITATION = ARXIV:1105.1779;%%
  %\cite{Calmet:2012cn}
%\bibitem{Calmet:2012cn} 
%  X.~Calmet, D.~Fragkakis and N.~Gausmann,
  ``Non thermal small black holes,'' chapter 8 in A.J. Bauer and D.G.Eiffel editors, Black Holes: Evolution, Theory and Thermodynamics Nova Publishers, New York, 2012,   arXiv:1201.4463 [hep-ph].
  %%CITATION = ARXIV:1201.4463;%%


%\cite{Calmet:2012fv}
\bibitem{Calmet:2012fv} 
  X.~Calmet and N.~Gausmann,
  %``Non-thermal quantum black holes with quantized masses,''
  Int.\ J.\ Mod.\ Phys.\ A {\bf 28}, 1350045 (2013)
  [arXiv:1209.4618 [hep-ph]].
  %%CITATION = ARXIV:1209.4618;%%
  %5 citations counted in INSPIRE as of 31 Oct 2014


 
   \bibitem{calmetbook2}
 X.~Calmet, ``Fundamental Physics with Black Holes,''
chapter 1 of ÒQuantum Aspects of Black Holes,Ó X. Calmet (Ed.), volume 178 of the series ÒFundamental Theories of Physics,Ó Springer, 2015, ISBN 978-3-319-10851-3. 

%\cite{Gingrich:2009hj}
\bibitem{Gingrich:2009hj} 
  D.~M.~Gingrich,
  %``Quantum black holes with charge, colour, and spin at the LHC,''
  J.\ Phys.\ G {\bf 37}, 105008 (2010)
  [arXiv:0912.0826 [hep-ph]].
  %%CITATION = ARXIV:0912.0826;%%
  %26 citations counted in INSPIRE as of 04 Dec 2014
  
     
 %\cite{Belyaev:2012qa}
\bibitem{Belyaev:2012qa} 
  A.~Belyaev, N.~D.~Christensen and A.~Pukhov,
  %``CalcHEP 3.4 for collider physics within and beyond the Standard Model,''
  Comput.\ Phys.\ Commun.\  {\bf 184}, 1729 (2013)
  [arXiv:1207.6082 [hep-ph]].
  %%CITATION = ARXIV:1207.6082;%%
  %126 citations counted in INSPIRE as of 27 May 2014 

\bibitem{hepmdb}
M. Bondarenko, A. Belyaev, L. Basso, E. Boos, V. Bunichev {\it et al.},
``High Energy Physics Model Database : Towards
decoding of the underlying theory" (within ``Les Houches 2011: Physics
at TeV Colliders New Physics Working Group Report"), arXiv:1203.1488
[hep-ph],
{https://hepmdb.soton.ac.uk}


  
 
  


 %\cite{Gingrich:2009da}
\bibitem{Gingrich:2009da} 
  D.~M.~Gingrich,
  %``Monte Carlo event generator for black hole production and decay in proton-proton collisions,''
  Comput.\ Phys.\ Commun.\  {\bf 181}, 1917 (2010)
  [arXiv:0911.5370 [hep-ph]].

   
 
 %\cite{Alwall:2006yp}
\bibitem{Alwall:2006yp} 
  J.~Alwall, A.~Ballestrero, P.~Bartalini, S.~Belov, E.~Boos, A.~Buckley, J.~M.~Butterworth and L.~Dudko {\it et al.},
  %``A Standard format for Les Houches event files,''
  Comput.\ Phys.\ Commun.\  {\bf 176}, 300 (2007)
  [hep-ph/0609017].
  %%CITATION = HEP-PH/0609017;%%
  %202 citations counted in INSPIRE as of 01 Dec 2014


%\cite{Pumplin:2002vw}
\bibitem{Pumplin:2002vw} 
  J.~Pumplin, D.~R.~Stump, J.~Huston, H.~L.~Lai, P.~M.~Nadolsky and W.~K.~Tung,
  %``New generation of parton distributions with uncertainties from global QCD analysis,''
  JHEP {\bf 0207}, 012 (2002)
  [hep-ph/0201195].
  %%CITATION = HEP-PH/0201195;%%
  %3896 citations counted in INSPIRE as of 27 May 2014
  
%\cite{Bityukov:1998zn}
\bibitem{Bityukov:1998zn} 
  S.~I.~Bityukov and N.~V.~Krasnikov,
  %``On observability of signal over background,''
   Nucl. Instrum. and Methods, {\bf A452}, pp 518-524 (2000)
  %2 citations counted in INSPIRE as of 03 Dec 2014
  
  %\cite{Aad:2014cka}
\bibitem{Aad:2014cka} 
  G.~Aad {\it et al.}  [ATLAS Collaboration],
  %``Search for high-mass dilepton resonances in pp collisions at sqrt(s) = 8 TeV with the ATLAS detector,''
  Phys.\ Rev.\ D {\bf 90}, 052005 (2014)
  [arXiv:1405.4123 [hep-ex]].
  %%CITATION = ARXIV:1405.4123;%%
  %27 citations counted in INSPIRE as of 07 Dec 2014

 
\end{thebibliography}
\end{document}